\documentclass[sigconf]{acmart}
\AtBeginDocument{%
  \providecommand\BibTeX{{%
    \normalfont B\kern-0.5em{\scshape i\kern-0.25em b}\kern-0.8em\TeX}}}

\usepackage{float}
\usepackage{newfloat}
\usepackage{multirow}
\usepackage{makecell}
\usepackage{xcolor}         
\usepackage{colortbl}
\usepackage{dsfont}
\usepackage{amsmath}
\usepackage{enumitem}

\usepackage[normalem]{ulem}
\useunder{\uline}{\ul}{}

\definecolor{myyellow}{rgb}{1,1, 0.6}
\definecolor{myorange}{rgb}{1, 0.8, 0.6}
\definecolor{myred}{rgb}{1, 0.6, 0.6}
\definecolor{second}{HTML}{FFDAB9}
\definecolor{best}{HTML}{FFC1C1}
\usepackage{multirow}
\usepackage{colortbl}
\usepackage[normalem]{ulem}
\usepackage{bbding}
\useunder{\uline}{\ul}{}
\usepackage{stackrel}
\usepackage{todonotes}
\usepackage{enumitem}
\usepackage{algpseudocode}
\usepackage[ruled,lined]{algorithm2e}
\usepackage[normalem]{ulem}
\usepackage{tcolorbox}

\renewcommand{\todo}[1]{\iffalse #1 \fi{\color{blue} \textbf{[TODO]}}}

\newcommand{\leftrarrows}{\mathrel{\raise.9ex\hbox{\oalign{%
  $\scriptstyle\leftarrow$\cr
  \vrule width0pt height.5ex$\hfil\scriptstyle\relbar$\cr}}}}
\newcommand{\lrightarrows}{\mathrel{\raise.9ex\hbox{\oalign{%
  $\scriptstyle\relbar$\hfil\cr
  $\scriptstyle\vrule width0pt height.5ex\smash\rightarrow$\cr}}}}
\newcommand{\Rrelbar}{\mathrel{\raise.9ex\hbox{\oalign{%
  $\scriptstyle\relbar$\cr
  \vrule width0pt height.5ex$\scriptstyle\relbar$}}}}

\newcommand{\longlongleftrightarrows}{\leftrarrows\joinrel\Rrelbar\joinrel\Rrelbar\joinrel\Rrelbar\joinrel\Rrelbar\joinrel\Rrelbar\joinrel\Rrelbar\joinrel\Rrelbar\joinrel\Rrelbar\joinrel\Rrelbar\joinrel\lrightarrows}

\DeclareMathOperator*{\argminA}{arg\,min}

\copyrightyear{2024}
\acmYear{2024}
\setcopyright{acmlicensed}\acmConference[MM '24]{Proceedings of the 32nd
ACM International Conference on Multimedia}{October 28-November 1,
2024}{Melbourne, VIC, Australia}
\acmBooktitle{Proceedings of the 32nd ACM International Conference on
Multimedia (MM '24), October 28-November 1, 2024, Melbourne, VIC, Australia}
\acmDOI{10.1145/3664647.3680574}
\acmISBN{979-8-4007-0686-8/24/10}

\begin{document}

\title{Semantic Codebook Learning for Dynamic Recommendation Models}

\author{Zheqi Lv}
\affiliation{%
  \institution{Zhejiang University}
  \city{Hangzhou}
  \country{China}}
\email{zheqilv@zju.edu.cn}

\author{Shaoxuan He}
\affiliation{%
  \institution{Zhejiang University}
  \city{Hangzhou}
  \country{China}}
\email{shxhe@zju.edu.cn}

\author{Tianyu Zhan}
\affiliation{%
  \institution{Zhejiang University}
  \city{Hangzhou}
  \country{China}}
\email{yuzt@zju.edu.cn}

\author{Shengyu Zhang}
\affiliation{%
  \institution{Zhejiang University}
  \city{Hangzhou}
  \country{China}\\
  \institution{Shanghai Institute for Advanced Study, Zhejiang University}
  \city{Shanghai}
  \country{China}}
\authornote{Corresponding authors.}
\email{sy_zhang@zju.edu.cn}

\author{Wenqiao Zhang}
\affiliation{%
  \institution{Zhejiang University}
  \city{Hangzhou}
  \country{China}}
\email{wenqiaozhang@zju.edu.cn}

\author{Jingyuan Chen}
\affiliation{%
  \institution{Zhejiang University}
  \city{Hangzhou}
  \country{China}}
\email{jingyuanchen@zju.edu.cn}

\author{Zhou Zhao}
\affiliation{%
  \institution{Zhejiang University}
  \city{Hangzhou}
  \country{China}}
\authornotemark[1]
\email{zhaozhou@zju.edu.cn}

\author{Fei Wu}
\affiliation{%
  \institution{Zhejiang University}
  \city{Hangzhou}
  \country{China}}
\email{wufei@zju.edu.cn}

\renewcommand{\shortauthors}{author name and author name, et al.}

\begin{abstract}
\label{sec:abstract}
Dynamic sequential recommendation (DSR) can generate model parameters based on user behavior to improve the personalization of sequential recommendation under various user preferences. However, it faces the challenges of large parameter search space and sparse and noisy user-item interactions, which reduces the applicability of the generated model parameters. 
The Semantic Codebook Learning for Dynamic Recommendation Models (SOLID) framework presents a significant advancement in DSR by effectively tackling these challenges. By transforming item sequences into semantic sequences and employing a dual parameter model, SOLID compresses the parameter generation search space and leverages homogeneity within the recommendation system. The introduction of the semantic metacode and semantic codebook, which stores disentangled item representations, ensures robust and accurate parameter generation. Extensive experiments demonstrates that SOLID consistently outperforms existing DSR, delivering more accurate, stable, and robust recommendations.
\end{abstract}



\maketitle
\section{Introduction}
\label{sec:introduction}

\begin{sloppypar}
Nowadays, as an important branch of recommendation systems, sequential recommendation has emerged, including DIN~\cite{ref:din}, GRU4Rec~\cite{ref:gru4rec}, SASRec~\cite{ref:sasrec}, BERT4Rec~\cite{ref:bert4rec} and other models that are crucial in the field of recommendation systems.
However, the behavior logic of most users is not universally applicable, and as interests can change, it necessitates that sequence recommendation models be able to adjust their parameters in real-time according to the user's current interest preferences. Consequently, dynamic sequential recommendation models (DSR) like DUET~\cite{ref:duet} and APG~\cite{ref:apg_rs1} have been developed.
\end{sloppypar}

\begin{figure*}[t]
\includegraphics[width=0.88\textwidth]{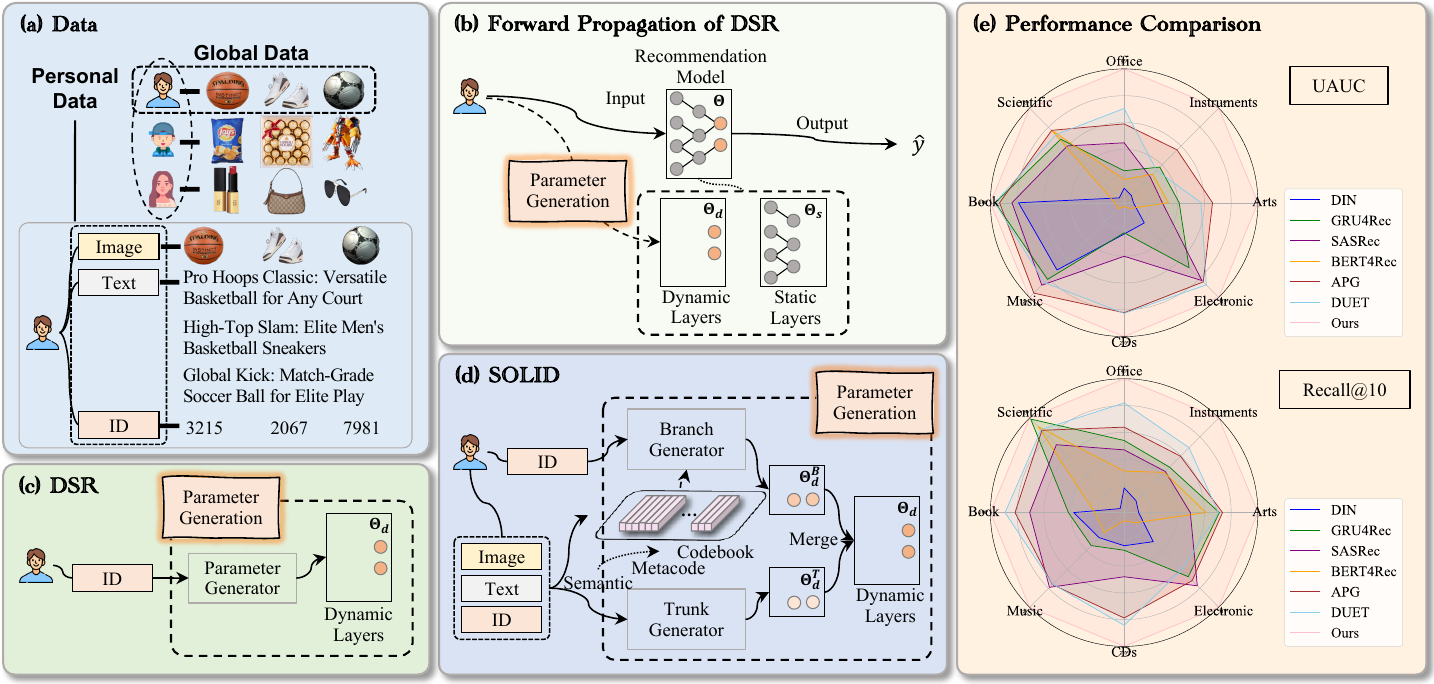}
\vspace{-0.3cm}
\centering\caption{ 
(a) describes multimodal user behavior data that includes images, text, and IDs.
(b) describes the forward propagation of DSR, which is divided into two pathways: the first pathway processes user behavior data composed of IDs through a parameter generator to produce the parameters for the dynamic layers of the primary model. The second pathway processes the same ID-based user behavior data through the primary model's static layer, then through the dynamic layer, resulting in the prediction output.
(c) and (d) compare the parameter generation patterns of existing DSR and SOLID.
(e) compares the performance of our method and SR models and DSR Models on four multi modal recommendation datasets and four single modal recommendation datasets. The results show that our method significantly enhances performance on extensive datasets.
}
\label{ref:introduction}
\vspace{-0.4cm}
\end{figure*}

The DSR paradigm consists of two parts: (1) The primary model. This model has a structure similar to conventional sequential recommendation models like SASRec, but it is divided into a static layer and a dynamic layer. The parameters of the static layer remain unchanged after pre-training, whereas the parameters of the dynamic layer change with the user's behavior. (2) The parameter generation model. This is mainly used to sparse user behavior and generate the parameters for the dynamic layer of the primary model based on this behavior. The DSR paradigm enables traditional static sequential recommendation models to quickly adjust their parameters according to the potential shift of interests and intentions reflected in user behaviors, thus dynamically obtaining more interest-aligned models in real time.

Despite the promising potential of Dynamic Sequential Recommendation (DSR) systems, they face significant challenges, primarily stemming from the item-to-parameter modeling scheme: (1) A large number of items result in a vast search space for the parameter generation model. Slight variations in user behavior sequences, such as "shirt, tie, suit" versus "tie, shirt, suit," which suggest similar preferences, can unpredictably alter the item-to-parameter modeling, introducing complexity and potential instability. (2) The interaction between users and items is generally sparse and potentially noisy (\textit{e.g.}, the notorious implicit feedback issue), leading to heterogeneous behavior sequences that complicate the learning of accurate item representations. This results in inaccurate item representation learning, weakening the precision of model parameter customization based on item sequence features, and further exacerbating the inaccuracy of generated parameters.

To address these issues, we propose the \textbf{S}emantic C\textbf{o}debook \textbf{L}earning for Dynam\textbf{i}c Recommen\textbf{d}ation Models (SOLID). The core objective of SOLID is to compress the search space of the parameter generation model, promoting homogeneity signals utilization within the recommendation system. We construct a semantic codebook that better utilizes these homogeneity signals. In the codebook, item representations are disentangled into semantics that are learned to be absorbed in the codebook elements, such that the homogeneity between items in the disentangled latent space can be established. The user-item interactions are transformed into density-enriched user-semantic interactions in the latent space. The enriched density reduces the heterogeneity and complexity of user behavior space modeling in the parameter generator. Moreover, SOLID shifts from a traditional item sequence-based parameter generation mode to a dual (item sequence + semantic sequence) → model parameter generation mode, effectively merging both uniform and diverse information in a structured manner. Uniform information derived from the semantic-to-parameter part is utilized to develop parameters that generalize across certain user behaviors, while diverse information allows for the crafting of specific parameters tailored to individual behavioral nuances. Crucially, by aligning the dimensions of the codebook with those of the semantic encoder, we transform the semantic encoder into a meta-code that serves as an initial state for the codebook, further easing the modeling of parameter generation.

Specifically, to reduce the search space of the parameter generation model through the semantic codebook, SOLID involves three main modules. Initially, SOLID employs a pretrained model to extract semantic components from item, image, and text features. This disentanglement transitions the focus from item sequences to semantic sequences, shifting the modeling approach from item-based to semantics-based parameter generation. This design results in trunk parameters that generalize behaviors from the entire user base to specific groups, and branch parameters that cater to individual user behaviors, both derived from semantic and item sequences respectively. Parameters derived from items are tightly controlled (e.g., ±0.01) before their integration into the dynamic layer of the primary model, ensuring a responsive and adaptive system based on real-time user activity. Despite this, branch parameters still adhere to an item-centric approach, necessitating the use of a Semantic Codebook (SC) to maintain personalization and stability in representation. This codebook stores semantic vectors of behavior, progressively aligned with the nearest matches during learning. The weights of the semantic encoder are used to initialize the SC, easing the semantic codebook learning. As shown in Figure~\ref{ref:introduction}, SOLID is designed to pursue the precision, stability, and clarity of model parameter generation, trying to promote the dynamic recommendation model's response to sparse, heterogeneous, and potentially noisy user behaviors.

Our contributions can be summarized as:
\begin{sloppypar}
\begin{itemize}[itemsep=1pt,topsep=2pt,leftmargin=20pt]
\item We pointed the limitations of the existing DSR paradigm and designed the SOLID framework to address these deficiencies.
\item We first learned to disentangle the parameter generation mode, which ensures that the generated model parameters contain both common and personalized knowledge.
\item We transformed the semantic encoder into a semantic metacode to enhance the semantic codebook learning.
\item We conducted extensive experiments on multiple datasets, which demonstrates the rationality and efficacy of SOLID.
\end{itemize}
\end{sloppypar}

\section{Related Work}
\label{sec:related_work}

\subsection{Sequential Recommendation}
Recommendation system predicts user preferences based on user behavior history~\cite{su2023enhancing,su2023personalized,zhang2021mining,zhang2022latent,lin2024temporally,lin2024data,li2023trustworthy,li2023should,xiao2023reconsidering,chen2021deep,zhang2021causal,zhang2023reformulating,liu2024learning,liao2023ppgencdr}. Sequential recommendation, as an important branch of the recommendation system, arranges users' recent historical behaviors in chronological order to more accurately capture users' recent preferences. Recent advancements~\cite{ref:gru4rec,ref:din,ref:sasrec,ref:bert4rec,ref:srgnn,ref:surge,lv2023parameters,ref:apg_rs1,ref:duet,liu2023joint} have shifted towards deep learning-based sequential recommendation systems. For instance, GRU4Rec~\cite{ref:gru4rec} employs Gated Recurrent Units to effectively model sequential behavior, demonstrating impressive results. Additionally, DIN~\cite{ref:din} and SASRec~\cite{ref:sasrec} incorporate attention mechanisms and transformers, respectively. BERT4Rec~\cite{ref:bert4rec} further applies BERT for superior outcomes in recommendation task. The models have significantly impacted academic research and industry practices. However, these SR Models struggle to achieve optimal performance across every data distribution when dealing with users' real-time changing behaviors and interest preferences. 

\subsection{Disentangled Representation Learning}
The goal of disentangled representation learning is to parse the data into distinct, interpretable components by identifying different underlying latent factors~\cite{bengio2013representation_dis1,cen2020controllable_dis2}. 
Variational autoencoders (VAE)~\cite{chen2022intent_dis3} and $\beta-$VAE~\cite{higgins2017beta_dis4} provide more possibilities for disentangled learning by adjusting the balance between the model's disentanglement ability and its ability to represent information.
By incorporating multi-interest methods~\cite{kingma2013auto_dis5,ma2020disentangled_dis6} along with disentangled representation learning, several studies~\cite{wang2022disentangled_dis7,wang2022disentangled_dis8,wang2021multimodal_dis9,wang2023curriculum_dis10,zhang2020content_dis11} have demonstrated significant advancements in recommendation tasks.
We draw on the idea of disentangling and apply it to dynamic model parameter generation to reduce the parameter search space and leverage the homogeneous information of user behavior.

\subsection{Dynamic Neural Network}
\begin{sloppypar}
Research in dynamic neural networks focuses on HyperNetworks~\cite{ref:hypernetworks} and Dynamic Filter Networks~\cite{ref:dfn}, which have better ability to adapt to distribution deviations than traditional static model learning or other efficient fine-tuning strategies~\cite{wang2024graph,wang2023deconfounded,chen2024learning,tang2024oodkd,huang2022fastdiff,huang2023make,li2022fine,zhang2021consensus,zhu2024mario,zhu2024graphcontrol,zheng2023rethinking,feng2023towards}. Similar situations also exist in the study of large models~\cite{zhu2024efficient,zhu2023bridging,zhu2024model,zhang2024hyperllava}. HyperNetworks, introduced by Ha et al.~\cite{ref:hypernetworks}, use one neural network to dynamically generate parameters for another, reducing the number of parameters needed and achieving model compression. This concept has led to extensive exploration and enhancements in various applications~\cite{ref:hypernetwork_continual_learning,ref:hypernetwork_graph,ref:hypernetwork_meta_learning,ref:hypernetwork_federated_learning,ref:hypernetwork_hyperstyle,ref:hypernetwork_hyperinverter,zhang2024revisiting,fu2024diet,zhang2024hyperllava,tang2024modelgpt}. Some recent research includes: HyperInverter~\cite{ref:hypernetwork_hyperinverter}, HyperStyle~\cite{ref:hypernetwork_hyperstyle}, Detective~\cite{zhang2024revisiting} introduces dynamic neural networks into multiple computer vision tasks to improve the model's personalization capabilities under various data distributions. IntellectReq~\cite{lv2024intelligent} detects when such dynamic networks need to modify parameters to adapt to samples, thereby achieving better performance with fewer parameter modifier calls. APG~\cite{ref:apg_rs1} and DUET~\cite{ref:duet} are the latest and state-of-the-art examples of using dynamic neural networks for sequence recommendation. However, existing DSR models are affected by the heterogeneity of user behavior, the sparsity of user-item interactions, etc., leading to drawbacks such as an overly large parameter search space and inaccurate parameter generation. Our method effectively addresses these shortcomings.
\end{sloppypar}
\section{Methodology}
\label{sec:method}
\begin{figure*}[t]
  \centering
  \includegraphics[width=0.9\linewidth]{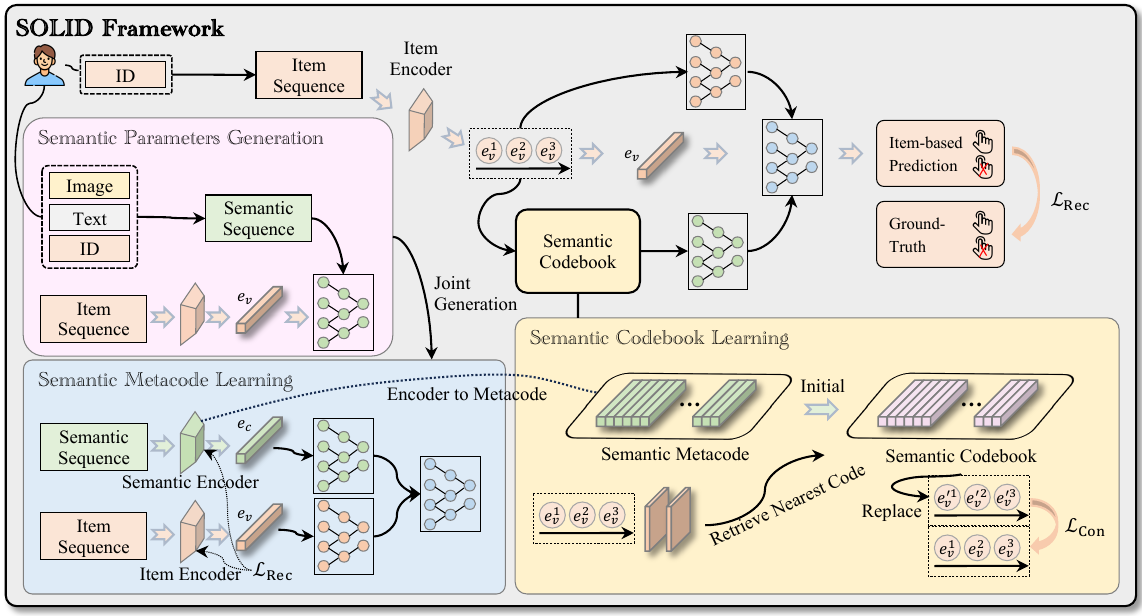}
  \vspace{-0.3cm}
  \caption{The framework of the SOLID, which consists of three main modules: Semantic Parameter Generation (SPG), Semantic Metacode Learning (SML), and Semantic Codebook Learning (SCL). SPG first converts item representations into semantics and constructs a semantic sequence to generate parameters in a structured manner. Subsequently, SML generates model parameters based on both the item sequence and the semantic sequence, and it jointly trains the model, accommodating both homogeneous and heterogeneous information. More importantly, the semantic encoder it learns can be transformed into metacode, which then provides a good initial value for the codebook. Finally, SCL learns a semantic codebook to improve the process of the parameter generation. Among them, $\mathcal{L}_{\text{Rec}}=l_{\text{CE}}(y, \hat{y}), \mathcal{L}_{\text{Con}} = l_{\text{MSE}}(\mathbf{E}_v, \mathbf{E}_v')$.}
  \label{fig:method}
\vspace{-0.4cm}
\end{figure*}
\subsection{Notations and Problem Formulation}
First, we introduce the notation in sequential recommendations. 

\subsubsection{Data.} We use $\mathcal{X_{\rm{ori}}}=\{u, v, s_v\}$ to represent a piece of data, $\mathcal{X_{\rm{dec}}}=\{u, c, s_c\}$ to represent a piece of disentangled data, $\mathcal{X_{\rm{mm}}}=\{i, t\}$ to represent multimodal information, and $\mathcal{Y}=\{y\}$ to represent the label indicating whether the user will interact with the item. In brief, $\mathcal{X}=\mathcal{X_{\rm{ori}}} \cup \mathcal{X_{\rm{dec}}} \cup \mathcal{X_{\rm{mm}}} =\{u, v, s_v, c, s_c, i, t\}$, where $u, v, c, s_v, s_c, i, t$ represent user ID, item ID, category ID, user's click sequence consists of item ID, user's click sequence consists of category ID, the image of the item, and the title of the item respectively. We represent the dataset as $\mathcal{D}$, where $\mathcal{D} = \{X, Y\}$. More specifically, we use $\mathcal{D}_{\rm{Train}}$ to represent the training set and $\mathcal{D}_{\rm{Test}}$ to represent the test set. Roughly speaking, let $\mathcal{L}$ be the loss obtained from training on dataset $\mathcal{D}_{\rm{Train}}$. For simplicity, we simplify the symbol $\mathcal{D}_{\rm{Train}}$ to $\mathcal{D}$. Then, the model parameters $W$ can be obtained through the optimization function $\argminA \mathcal{L}$. The sequence length inputted into the model is set to \(L_s\), so the lengths of both \(s_v\) and \(s_c\) in a sample are \(L_s\).

\subsubsection{Model.} The recommendation model is represented by $\mathcal{M}$ and the parameters of the $\mathcal{M}$ is $\Theta$, where $\Theta={\Theta_s, \Theta_d}$. The model $\mathcal{M}_v$ is utilized to generate the $\Theta_d$ according to the item id sequence $s_v$, $\mathcal{M}_c$ is utilized to generate the $\Theta_d$ according to the category id sequence $s_c$,  $\mathcal{M}(\cdot)$ and $\mathcal{M}_v(\cdot)$ represent the forward propagation processes of two models, where $\cdot$ denotes the input.

\subsubsection{Feature.} We use $\mathbf{E_v}$ and $\mathbf{E_c}$ to represent the item feature set and semantic feature set extracted from $s_v$ and $s_c$ respectively. Specifically, $\mathbf{E_v} = \{e_v^1, e_v^2, ..., e_v^{L_s}\}$, $\mathbf{E_c} = \{e_c^1, e_c^2, ..., e_c^{L_s}\}$. \(\mathbf{e_v}\) and \(\mathbf{e_c}\) are the sequence features obtained through sequence feature extraction models such as Transformer or GRU, via \(\mathbf{E_v}\) and \(\mathbf{E_c}\), respectively.
The length of an item representation or a semantic representation is set to $L_r$.

\subsubsection{Formula.} Sequential Recommendation Models (SR), Dynamic Sequential Recommendation Models (DSR), and Disentangled Multimodal Dynamic Sequential Recommendation Models (SOLID) can be formalized as follows:
\begin{equation}
\resizebox{0.41\textwidth}{!}{
$
\textbf{SR}:
\underbrace{\mathcal{M}(\mathcal{X}_{\rm{ori}};\Theta)}_{\rm{Recommendation\ Procedure}} \stackrel[\text{Output}]{\text{Gradients}}{\longlongleftrightarrows} \underbrace{(\hat{\mathcal{Y}} {\Longleftrightarrow} \mathcal{Y})}_{\text{Loss Calculation}}.
$
}
 \label{eq:problem_formulation_sr}
\end{equation}

\begin{equation}
\resizebox{0.44\textwidth}{!}{
$
\textbf{DSR}:
\underbrace{\mathcal{M}(\mathcal{X}_{\rm{ori}};\Theta_s, \Theta_d=\mathcal{M}_v(\mathcal{X}_{\rm{ori}}))}_{\rm{Recommendation\ Procedure}} \stackrel[\text{Output}]{\text{Gradients}}{\longlongleftrightarrows} \underbrace{(\hat{\mathcal{Y}} {\Longleftrightarrow} \mathcal{Y})}_{\text{Loss\ Calculation}}.
$
}
\label{eq:problem_formulation_dsr}
\end{equation}

\begin{equation}
\resizebox{0.47\textwidth}{!}{
$
\textbf{SOLID}:
\left
\{\begin{aligned}
& \mathcal{X}_{\text{ori}}, \mathcal{X}_{\text{mm}} \mapsto c = f(v, i, t) \mapsto \mathcal{X}_{\text{dec}}, \\
& \Theta_d=\mathcal{M}_v(\mathcal{X}_{\rm{ori}}) \oplus \mathcal{M}_c(\mathcal{X}_{\rm{dec}}),\\
& \underbrace{\mathcal{M}(\mathcal{X}_{\rm{ori}};\Theta_s, \Theta_d)}_{\rm{Recommendation\ Procedure}} \stackrel[\text{Output}]{\text{Gradients}}{\longlongleftrightarrows} \underbrace{(\hat{\mathcal{Y}} {\Longleftrightarrow} \mathcal{Y})}_{\text{Loss\ Calculation}}.
\end{aligned}
\right.
$
}
 \label{eq:problem_formulation_dmdsr}
\end{equation}
In the aforementioned formula, $a \rightarrow b$ indicates indicates information transfer from $a$ to $b$, with the text next to it representing the content of the transfer. $ a \mapsto b $ signifies that $ b $ is derived from $ a $.

\subsection{Preliminary}
\subsubsection{Sequential Recommendation Models}
Here we first retrospect the paradigm of sequential recommendation. 

In the training stage, the loss can be calculated to optimize the sequential recommendation models as follows,
\begin{equation}
    \mathop{\rm min}_{\Theta}\mathcal{L}=\sum\nolimits_{u, v, s_v, y\in\mathcal{D}}l_{\text{CE}}(y, \hat{y}=\mathcal{M}(u, v, s_v; \Theta)).
\label{eq:loss_general}
\end{equation}
The loss function can set to CE (Cross Entropy) loss and MSE (Mean Squared Error) loss, etc. However, since sequential recommendation often focuses more on CTR (Click-Through Rate) prediction tasks, and this paper is also focused on CTR prediction, the recommendation loss in this paper is CE loss and represented by \(l_{\text{CE}}\).

\subsubsection{Dynamic Sequential Recommendation Models}
DSR generate model parameters based on users' real-time user behaviors. Then the updated model is used for current recommendations. In this paper, the network layer that can adjust model parameters as the data distribution changes is called an adaptive layer.

\begin{sloppypar} 
DSR treat the parameters of one of the adaptive layers as a matrix $K \in \mathbb{R}^{N_{in}\times N_{out}}$, where $N_{in}$ and $N_{out}$ represent the number of input neurons and output neurons of a fully connected layer (FCL), respectively. 
DSR utilize a encoder $E_v$ to extract the sequence feature $\boldsymbol{e}_{v}$ from the user's behavior sequence $s_v$ to generate the parameters of the model's adaptive layers.
\end{sloppypar}
\begin{equation}
\label{eq:lightweight_encoder}
    \theta_d = \mathcal{M}_v(E_v (s_v)), 
\end{equation}
After parameter generation, the parameters of the model will be reshaped into the shape of $K$.

During training, all layers of the $\mathcal{M}_v$ are optimized together with the static layers of the $\mathcal{M}$. The loss function $\mathcal{L}$ is defined as follows:
\begin{equation}
    \label{eq:loss_func}
\mathop{\rm min}_{\Theta_s, \Theta_v}\mathcal{L}=\sum\nolimits_{u, v, s_v, y\in\mathcal{D}}l_{\text{CE}}(y, \hat{y} = \mathcal{M}(u,v,s_v;\Theta_s, \Theta_d)).
\end{equation}
Although the Item-based Dynamic Recommendation Model can obtain personalized model parameters based on users' real-time behavior and achieve superior performance, it also faces multiple challenges. 1) The user-item interaction is extremely sparse, leading to inaccurate item representation learning, making the model parameters customized based on item-based features inaccurate. 2) The personalized model parameters obtained by this strategy are highly mixed. 3) The generated parameters are not subject to any constraints, which poses challenges to the stability of the generated model. So we design the novel methods to address the challenges mentioned above.

\subsection{SOLID Framework}
The architecture of our proposed SOLID is shown in the Figure~\ref{fig:method}.
\subsubsection{Semantic Parameter Generation}
Transforming the Item-based Dynamic Recommendation Model into a Semantic-based Dynamic Recommendation Model is an important step in disentangling personalized model parameters.
First, items need to be transformed into semantics. For data without category labels, clustering can be directly applied to obtain semantics, i.e., 
\begin{equation}
\text{Cluster}(\{e_i\}_{i=1}^{\mathcal{N}}) \mapsto \{c_i\}_{i=1}^{\mathcal{N}}, c_i \in \{1, 2, ..., k\}.
\end{equation}
For data with category labels, since the same item often belongs to multiple categories, we select a primary category as semantic it. First, we define the centroid $m_c$ of each category $c$, which is the average of embeddings $e$ for all items belonging to category $c$. Assuming $n_c$ is the number of items belonging to category $c$, the centroid $m_c$ for category $c$ can be represented as:

\begin{equation}
m_c = \frac{1}{n_c} \sum\nolimits_{v \in c} (e_v \text{ or } e_i \text{ or } e_t),
\end{equation}
where $e_v, e_i, e_t$ are the representation of item ID $v$, item image $i$, item title $t$, respectively. Next, we compute its distance to each category center $m_c$. Assuming we use the Euclidean distance, it can be represented as,
\begin{equation}
d(v, c) = \| (e_v \text{ or } e_i \text{ or } e_t) - m_c \|,
\end{equation}
where $\| \cdot \|$ denotes the norm of the vector, typically the Euclidean norm. Finally, we select the closest category as the semantic for item $v$. That is, the semantic $c_p$ for item $v$ can be represented as:

\begin{equation}
c_p = \arg\min_{c} d((v \text{ or } i \text{ or } t), c).
\end{equation}

After converting items into semantics, a semantic-to-parameter model can be trained. The training process is similar to that of the item-to-parameter model. The only differences are that the input for the item-to-parameter model is an item sequence, whereas for the semantic-to-parameter model, it is a semantic sequence; similarly, the outputs are the target item and target semantic, respectively.
\begin{equation}
\left
\{\begin{aligned}
\mathop{\rm min}_{\Theta_s, \Theta_c}\mathcal{L} &= \sum\nolimits_{u, v, s_c, y\in\mathcal{D}}l_{\text{CE}}(y, \hat{y}), \\
\hat{y} &= \mathcal{M}(u,v,s_c;\Theta_s, \Theta_d), \\
\Theta_d &= \mathcal{M}_c(E_c(s_c)).
\end{aligned}
\right.
\end{equation}
In the above equation, $E_c$ represents the semantic encoder, which is similar to the item encoder $E_v$.

\subsubsection{Semantic Metacode Learning}
To balance the use of personalized user behavior information and homogeneous information from similar user behaviors, we combine the item-to-parameter and semantic-to-parameter models for the parameter generation process. The former's advantage lies in providing personalized information, but its disadvantage is the inaccuracy in parameter generation due to strong data heterogeneity and sparse user-item interactions. The latter's advantage is providing homogeneous information from similar user behaviors, and dense user-item interactions make the parameter generation process more robust. However, its disadvantage is that the semantic sequence is less personalized compared to the item sequence.

Therefore, our approach primarily uses the semantic-to-parameter method to generate the main part of the model parameters. Since similar semantic sequences are easier to obtain than similar item sequences, the parameters derived from the semantic sequence can be viewed as a user group model. Then, the item-to-parameter method is used as a branch, with parameters generated from item sequences being constrained within a smaller threshold and merged with the parameters obtained from the semantic sequence. This merging process is seen as a transition from a user group model to an individual user model, thus balancing homogeneous information and personalized information. Therefore, the training process can be formulated as the following optimization problem,
\begin{equation}
\label{eq:metacode_learning}
\left
\{\begin{aligned}
\mathop{\rm min}_{\Theta_s, \Theta_c, \Theta_v}\mathcal{L} &= \sum\nolimits_{u, v, s_c, y\in\mathcal{D}}l_{\text{CE}}(y, \hat{y}), \\
\hat{y} &= \mathcal{M}(u,v,s_v;\Theta_s, \Theta_d), \\
\Theta_d &= \mathcal{M}_c(E_c(s_c)) + {\rm{Clip}}(\mathcal{M}_v(E_v(s_v))); \mathcal{T}),
\end{aligned}
\right.
\end{equation}
where $\mathcal{T}$ is a hyperparameter used to control the threshold for parameter deviation, thereby also controlling the impact of personalized information on the model parameters. Semantic Encoder can be transformed into a Semantic Metacode(SM), which can be used to further enhance the initialization of the Semantic Codebook for the item-to-parameter process. The Semantic Metacode can be effectively learned through the above process.

\subsubsection{Semantic Codebook Learning}
Even if the model parameter generation process is disentangled, the item-to-parameter mode is still needed because it is the source of personalized information. Therefore, to further improve the accuracy of the item-to-parameter mapping, we design a Semantic Codebook (SC). 
Upon obtaining the semantic metacode, we initialize the semantic codebook with it. Subsequently, we continue using the trunk and branch method of parameter generation, specifically semantic-to-parameter and item-to-parameter, to derive the parameters for the adaptive layer of the model. In the branch branch, the item representations are replaced with semantic codes from the codebook, which are then used to further predict model parameters. The generated model parameters are used for click prediction on item sequences, just as before, ultimately allowing for the training of the semantic codebook. The specific method for computing the loss is described below.
SC is denoted as \(D\), and \(D \in \mathbb{R}^{\mathcal{N}_c \times L_r} \). Specifically, we first use the weights of the semantic encoder in the semantic-to-parameter to initialize the item representation, as their dimensions are the same. Then, we encode the user's item representation. For a piece of data, as introduced in the notation description section, its item representation is \(\mathbf{E}_v = \{e_v^1, e_v^2, ..., e_v^{L_s}\}\). Afterward, we find the closest feature in the SC to replace each item representation in the set \(\mathbf{E}_v\), obtaining \(\mathbf{E}_v' = \{e_v'^1, e_v'^2, ..., e_v'^{L_s}\}\), and the sequence feature obtained from \(\mathbf{E}_v'\) is \(e_v'\).
Subsequently, we compute the MSE loss between the item representation set \(\mathbf{E}_v'\) obtained from the SC and the original set \(\mathbf{E}_v\), and incorporate it into the training process as follows,
\begin{equation}
\label{eq:codebook_learning}
\left
\{\begin{aligned}
\mathop{\rm min}_{\Theta_s, \Theta_c, \Theta_v}\mathcal{L} &= \sum\nolimits_{u, v, s_c, y\in\mathcal{D}}l_{\text{CE}}(y, \hat{y}) + \lambda l_{\text{MSE}}(\mathbf{E}_v, \mathbf{E}_v'), \\
\hat{y} &= \mathcal{M}(u,v,s_v;\Theta_s, \Theta_d), \\
\Theta_d &= \mathcal{M}_c(e_c) + {\rm{Clip}}(\mathcal{M}_v(e_v')); \mathcal{T}),
\end{aligned}
\right.
\end{equation}
where $l_{\text{MSE}}$ represents the MSE loss, and the $\lambda$ is a hyperparameter.

\subsubsection{Pseudo Code of SOLID}
\label{sec:pseudo_code}
Algorithm~\ref{alg:pseudo_code} shows the pseudo code of SOLID. $(x)$ represents that $x$ is a intermediate variable.

\renewcommand{\algorithmicrequire}{\textbf{Input:}} 
\renewcommand{\algorithmicrequire}{\textbf{Initialization:}}

\begin{algorithm}[!h]
\begin{flushleft}
  \caption{Pseudo code of SOLID}
    
\textbf{Module 1:}~\colorbox{gray!30}{$\rhd$~\emph{Item to Semantic}}
    \resizebox{0.46\textwidth}{!}{
    \begin{tcolorbox}[sharp corners, colframe=gray!80!white, colback=white, boxrule=0.5mm, left=0pt, right=0pt, top=-3pt, bottom=-4pt, boxsep=5pt]
    \begin{algorithmic}
    \State \textbf{Target}: Item Sequence $s_v$ $\mapsto$ Semantic Sequence $s_c$ 
    \State \textbf{Input}: Item Sequence $s_v$
    \State \textbf{Output}: Semantic Sequence $s_c$
    \end{algorithmic}
    \end{tcolorbox}
    }

\textbf{Module 2:}~\colorbox{gray!30}{$\rhd$~\emph{Semantic Parameter Generation}}
\resizebox{0.46\textwidth}{!}{
    \begin{tcolorbox}[sharp corners, colframe=gray!80!white, colback=white, boxrule=0.5mm, left=0pt, right=0pt, top=-3pt, bottom=-4pt, boxsep=5pt]
    \begin{algorithmic}
    \State \textbf{Target}: Semantic Sequence $s_c$ $\mapsto$ Semantic Parameter Generator $\mathcal{M}_c$ and Semantic Encoder $E_c$ 
    \State \textbf{Input}: Semantic Sequence $s_c$
    \State \textbf{Output}: (Parameter $\Theta_d$), Prediction $\hat{y}$
    \end{algorithmic}
    \end{tcolorbox}
    }

\textbf{Module 3:}~\colorbox{gray!30}{$\rhd$~\emph{Semantic Metacode Learning}}
\resizebox{0.46\textwidth}{!}{
    \begin{tcolorbox}[sharp corners, colframe=gray!80!white, colback=white, boxrule=0.5mm, left=0pt, right=0pt, top=-3pt, bottom=-4pt, boxsep=5pt]
    \begin{algorithmic}
    \State \textbf{Target}: Item Sequence $s_v$, Semantic Sequence $s_c$ $\mapsto$ Item Parameter Generator $\mathcal{M}_v$, Item Encoder $E_v$, Semantic Parameter Generator $\mathcal{M}_c$, and Semantic Encoder $E_c$ 
    \State \textbf{Input}: Item Sequence $s_v$, Semantic Sequence $s_c$
    \State \textbf{Output}: (Parameter $\Theta_d$), Prediction $\hat{y}$
    \end{algorithmic}
    \end{tcolorbox}
    }

\textbf{Module 4:}~\colorbox{gray!30}{$\rhd$~\emph{Semantic Codebook Learning}}
\resizebox{0.46\textwidth}{!}{
    \begin{tcolorbox}[sharp corners, colframe=gray!80!white, colback=white, boxrule=0.5mm, left=0pt, right=0pt, top=-3pt, bottom=-4pt, boxsep=5pt]
    \begin{algorithmic}
    \State \textbf{Target}: Item Sequence $s_v$, Semantic Sequence $s_c$, Semantic Encoder $E_c$ $\mapsto$ Codebook $D$
    \State \textbf{Input}: Item Sequence $s_v$, Semantic Sequence $s_c$, (Semantic Encoder $E_c$)
    \State \textbf{Output}: (Parameter $\Theta_d$), Prediction $\hat{y}$
    \end{algorithmic}
    \end{tcolorbox}
    }

\textbf{Overview:}~\colorbox{gray!30}{$\rhd$~\emph{Training Procedure}}\\
    \textbf{Input}: Item Sequence $s_v$, Semantic Sequence $s_c$. \\
    \textbf{Output}: (Parameters $\Theta_d$), Prediction $\hat{y}$.\\
    \textbf{Initialization}: Randomly initialize the models $\mathcal{M}$, $\mathcal{M}_c$, $\mathcal{M}_v$ with parameters $\Theta_s$, $\Theta_c$, $\Theta_v$ respectively. \\
    Item Sequence $s_v$  $\mapsto$ Semantic Sequence $s_c$\\
    \Repeat {Convergence}{
     \If{$\mathcal{M}_c$ and $E_c$ have not yet been well-trained} 
     {
           Train as Eq.\ref{eq:metacode_learning} \\
           }
          }
  \textbf{Initialization}: Initialize $D$ via pretrained $E_c$\\
  \Repeat {Convergence}{
  \If{$\mathcal{M}_c$ and $E_c$ have not yet been well-trained} 
    {
        Train as Eq.\ref{eq:codebook_learning} \\
    }
    }
    \Return{$\mathcal{M}_c$, $\mathcal{M}_v$, $D$}.\\
\label{alg:pseudo_code}
\end{flushleft}
\end{algorithm}

\begin{table*}[ht]
\vspace{-0.1cm}
\caption{Performance comparison of the proposed method and baselines. The \textbf{best} results is in bold. 
}
\label{tab:main}
\vspace{-0.3cm}
\renewcommand{\arraystretch}{1.02}
\resizebox{1\textwidth}{!}{
\begin{tabular}{c|c|c|c|c|c|c|c|c|c|c|c|c|c|c|c}
\toprule[2pt]
\multicolumn{8}{c|}{\textbf{\texttt{Arts}}} & \multicolumn{8}{c}{\textbf{\texttt{Instruments}}} \\
\midrule
 &  & \multicolumn{6}{c|}{\textbf{Metrics}} &  &  & \multicolumn{6}{c}{\textbf{Metrics}} \\ 
 \cline{3-8} \cline{11-16}
\multirow{-2}{*}{\textbf{SR Model}} & \multirow{-2}{*}{\textbf{DSR Model}} & AUC & UAUC & NDCG@10 & Recall@10 & NDCG@20 & Recall@20 & \multirow{-2}{*}{\textbf{SR Model}} & \multirow{-2}{*}{\textbf{DSR Model}} & AUC & UAUC & NDCG@10 & Recall@10 & NDCG@20 & Recall@20 \\
\midrule \midrule
 & - & 0.8193 & 0.7559 & 0.2646 & 0.4696 & 0.2993 & 0.6054 &  & - & 0.7974 & 0.7463 & 0.2620 & 0.4576 & 0.2966 & 0.5991 \\
 & \cellcolor[HTML]{F2F2F2}APG & \cellcolor[HTML]{F2F2F2}0.8432 & \cellcolor[HTML]{F2F2F2}0.7786 & \cellcolor[HTML]{F2F2F2}0.2868 & \cellcolor[HTML]{F2F2F2}0.5024 & \cellcolor[HTML]{F2F2F2}0.3221 & \cellcolor[HTML]{F2F2F2}0.6363 &  & \cellcolor[HTML]{F2F2F2}APG & \cellcolor[HTML]{F2F2F2}0.8183 & \cellcolor[HTML]{F2F2F2}0.7534 & \cellcolor[HTML]{F2F2F2}0.2680 & \cellcolor[HTML]{F2F2F2}0.4606 & \cellcolor[HTML]{F2F2F2}0.3025 & \cellcolor[HTML]{F2F2F2}0.5962 \\
 & \cellcolor[HTML]{F2F2F2}Ours (APG) & \cellcolor[HTML]{F2F2F2}\textbf{0.8459} & \cellcolor[HTML]{F2F2F2}\textbf{0.7873} & \cellcolor[HTML]{F2F2F2}\textbf{0.2907} & \cellcolor[HTML]{F2F2F2}\textbf{0.5144} & \cellcolor[HTML]{F2F2F2}\textbf{0.3271} & \cellcolor[HTML]{F2F2F2}\textbf{0.6529} &  & \cellcolor[HTML]{F2F2F2}Ours (APG) & \cellcolor[HTML]{F2F2F2}\textbf{0.8274} & \cellcolor[HTML]{F2F2F2}\textbf{0.7769} & \cellcolor[HTML]{F2F2F2}\textbf{0.2918} & \cellcolor[HTML]{F2F2F2}\textbf{0.5006} & \cellcolor[HTML]{F2F2F2}\textbf{0.3257} & \cellcolor[HTML]{F2F2F2}\textbf{0.6364} \\
 & \cellcolor[HTML]{E7E6E6}DUET & \cellcolor[HTML]{E7E6E6}0.8338 & \cellcolor[HTML]{E7E6E6}0.7647 & \cellcolor[HTML]{E7E6E6}0.2837 & \cellcolor[HTML]{E7E6E6}0.4893 & \cellcolor[HTML]{E7E6E6}0.3185 & \cellcolor[HTML]{E7E6E6}0.6202 &  & \cellcolor[HTML]{E7E6E6}DUET & \cellcolor[HTML]{E7E6E6}0.8126 & \cellcolor[HTML]{E7E6E6}0.7499 & \cellcolor[HTML]{E7E6E6}0.2727 & \cellcolor[HTML]{E7E6E6}0.4658 & \cellcolor[HTML]{E7E6E6}0.3060 & \cellcolor[HTML]{E7E6E6}0.5970 \\
\multirow{-5}{*}{DIN} & \cellcolor[HTML]{E7E6E6}Ours (DUET) & \cellcolor[HTML]{E7E6E6}\textbf{0.8426} & \cellcolor[HTML]{E7E6E6}\textbf{0.7830} & \cellcolor[HTML]{E7E6E6}\textbf{0.3014} & \cellcolor[HTML]{E7E6E6}\textbf{0.5162} & \cellcolor[HTML]{E7E6E6}\textbf{0.3363} & \cellcolor[HTML]{E7E6E6}\textbf{0.6486} & \multirow{-5}{*}{DIN} & \cellcolor[HTML]{E7E6E6}Ours (DUET) & \cellcolor[HTML]{E7E6E6}\textbf{0.8207} & \cellcolor[HTML]{E7E6E6}\textbf{0.7613} & \cellcolor[HTML]{E7E6E6}\textbf{0.2850} & \cellcolor[HTML]{E7E6E6}\textbf{0.4885} & \cellcolor[HTML]{E7E6E6}\textbf{0.3183} & \cellcolor[HTML]{E7E6E6}\textbf{0.6181} \\
\midrule
 & - & 0.8434 & 0.7837 & 0.2799 & 0.4943 & 0.3169 & 0.6380 &  & - & 0.8103 & 0.7604 & 0.2770 & 0.4772 & 0.3102 & 0.6105 \\
 & \cellcolor[HTML]{F2F2F2}APG & \cellcolor[HTML]{F2F2F2}0.8416 & \cellcolor[HTML]{F2F2F2}0.7796 & \cellcolor[HTML]{F2F2F2}0.2828 & \cellcolor[HTML]{F2F2F2}0.4986 & \cellcolor[HTML]{F2F2F2}0.3196 & \cellcolor[HTML]{F2F2F2}0.6403 &  & \cellcolor[HTML]{F2F2F2}APG & \cellcolor[HTML]{F2F2F2}0.8171 & \cellcolor[HTML]{F2F2F2}0.7578 & \cellcolor[HTML]{F2F2F2}0.2746 & \cellcolor[HTML]{F2F2F2}0.4716 & \cellcolor[HTML]{F2F2F2}0.3089 & \cellcolor[HTML]{F2F2F2}0.6069 \\
 & \cellcolor[HTML]{F2F2F2}Ours (APG) & \cellcolor[HTML]{F2F2F2}\textbf{0.8463} & \cellcolor[HTML]{F2F2F2}\textbf{0.7897} & \cellcolor[HTML]{F2F2F2}\textbf{0.3023} & \cellcolor[HTML]{F2F2F2}\textbf{0.5242} & \cellcolor[HTML]{F2F2F2}\textbf{0.3378} & \cellcolor[HTML]{F2F2F2}\textbf{0.6589} &  & \cellcolor[HTML]{F2F2F2}Ours (APG) & \cellcolor[HTML]{F2F2F2}\textbf{0.8296} & \cellcolor[HTML]{F2F2F2}\textbf{0.7752} & \cellcolor[HTML]{F2F2F2}\textbf{0.2911} & \cellcolor[HTML]{F2F2F2}\textbf{0.4971} & \cellcolor[HTML]{F2F2F2}\textbf{0.3265} & \cellcolor[HTML]{F2F2F2}\textbf{0.6360} \\
 & \cellcolor[HTML]{E7E6E6}DUET & \cellcolor[HTML]{E7E6E6}0.8463 & \cellcolor[HTML]{E7E6E6}0.7809 & \cellcolor[HTML]{E7E6E6}0.2911 & \cellcolor[HTML]{E7E6E6}0.5061 & \cellcolor[HTML]{E7E6E6}0.3277 & \cellcolor[HTML]{E7E6E6}0.6430 &  & \cellcolor[HTML]{E7E6E6}DUET & \cellcolor[HTML]{E7E6E6}0.8236 & \cellcolor[HTML]{E7E6E6}0.7568 & \cellcolor[HTML]{E7E6E6}0.2699 & \cellcolor[HTML]{E7E6E6}0.4655 & \cellcolor[HTML]{E7E6E6}0.3058 & \cellcolor[HTML]{E7E6E6}0.6059 \\
\multirow{-5}{*}{GRU4Rec} & \cellcolor[HTML]{E7E6E6}Ours (DUET) & \cellcolor[HTML]{E7E6E6}\textbf{0.8466} & \cellcolor[HTML]{E7E6E6}\textbf{0.7915} & \cellcolor[HTML]{E7E6E6}\textbf{0.3111} & \cellcolor[HTML]{E7E6E6}\textbf{0.5368} & \cellcolor[HTML]{E7E6E6}\textbf{0.3460} & \cellcolor[HTML]{E7E6E6}\textbf{0.6694} & \multirow{-5}{*}{GRU4Rec} & \cellcolor[HTML]{E7E6E6}Ours (DUET) & \cellcolor[HTML]{E7E6E6}\textbf{0.8261} & \cellcolor[HTML]{E7E6E6}\textbf{0.7740} & \cellcolor[HTML]{E7E6E6}\textbf{0.2958} & \cellcolor[HTML]{E7E6E6}\textbf{0.4987} & \cellcolor[HTML]{E7E6E6}\textbf{0.3313} & \cellcolor[HTML]{E7E6E6}\textbf{0.6401} \\
\midrule
 & - & 0.8383 & 0.7737 & 0.2758 & 0.4852 & 0.3127 & 0.6273 &  & - & 0.8201 & 0.7586 & 0.2729 & 0.4705 & 0.3071 & 0.6051 \\
 & \cellcolor[HTML]{F2F2F2}APG & \cellcolor[HTML]{F2F2F2}0.8370 & \cellcolor[HTML]{F2F2F2}0.7687 & \cellcolor[HTML]{F2F2F2}0.2816 & \cellcolor[HTML]{F2F2F2}0.4884 & \cellcolor[HTML]{F2F2F2}0.3166 & \cellcolor[HTML]{F2F2F2}0.6222 &  & \cellcolor[HTML]{F2F2F2}APG & \cellcolor[HTML]{F2F2F2}0.8200 & \cellcolor[HTML]{F2F2F2}0.7523 & \cellcolor[HTML]{F2F2F2}0.2663 & \cellcolor[HTML]{F2F2F2}0.4601 & \cellcolor[HTML]{F2F2F2}0.3010 & \cellcolor[HTML]{F2F2F2}0.5929 \\
 & \cellcolor[HTML]{F2F2F2}Ours (APG) & \cellcolor[HTML]{F2F2F2}\textbf{0.8414} & \cellcolor[HTML]{F2F2F2}\textbf{0.7820} & \cellcolor[HTML]{F2F2F2}\textbf{0.3018} & \cellcolor[HTML]{F2F2F2}\textbf{0.5145} & \cellcolor[HTML]{F2F2F2}\textbf{0.3365} & \cellcolor[HTML]{F2F2F2}\textbf{0.6468} &  & \cellcolor[HTML]{F2F2F2}Ours (APG) & \cellcolor[HTML]{F2F2F2}\textbf{0.8234} & \cellcolor[HTML]{F2F2F2}\textbf{0.7573} & \cellcolor[HTML]{F2F2F2}\textbf{0.2699} & \cellcolor[HTML]{F2F2F2}\textbf{0.4622} & \cellcolor[HTML]{F2F2F2}\textbf{0.3065} & \cellcolor[HTML]{F2F2F2}\textbf{0.6029} \\
 & \cellcolor[HTML]{E7E6E6}DUET & \cellcolor[HTML]{E7E6E6}0.8345 & \cellcolor[HTML]{E7E6E6}0.7660 & \cellcolor[HTML]{E7E6E6}0.2727 & \cellcolor[HTML]{E7E6E6}0.4763 & \cellcolor[HTML]{E7E6E6}0.3101 & \cellcolor[HTML]{E7E6E6}0.6177 &  & \cellcolor[HTML]{E7E6E6}DUET & \cellcolor[HTML]{E7E6E6}0.8241 & \cellcolor[HTML]{E7E6E6}0.7599 & \cellcolor[HTML]{E7E6E6}0.2768 & \cellcolor[HTML]{E7E6E6}0.4760 & \cellcolor[HTML]{E7E6E6}0.3105 & \cellcolor[HTML]{E7E6E6}0.6076 \\
\multirow{-5}{*}{SASRec} & \cellcolor[HTML]{E7E6E6}Ours (DUET) & \cellcolor[HTML]{E7E6E6}\textbf{0.8469} & \cellcolor[HTML]{E7E6E6}\textbf{0.7867} & \cellcolor[HTML]{E7E6E6}\textbf{0.3022} & \cellcolor[HTML]{E7E6E6}\textbf{0.5216} & \cellcolor[HTML]{E7E6E6}\textbf{0.3382} & \cellcolor[HTML]{E7E6E6}\textbf{0.6560} & \multirow{-5}{*}{SASRec} & \cellcolor[HTML]{E7E6E6}Ours (DUET) & \cellcolor[HTML]{E7E6E6}\textbf{0.8270} & \cellcolor[HTML]{E7E6E6}\textbf{0.7661} & \cellcolor[HTML]{E7E6E6}\textbf{0.2843} & \cellcolor[HTML]{E7E6E6}\textbf{0.4827} & \cellcolor[HTML]{E7E6E6}\textbf{0.3198} & \cellcolor[HTML]{E7E6E6}\textbf{0.6206} \\
\midrule
 & - & 0.8322 & 0.7791 & 0.2752 & 0.4885 & 0.3126 & 0.6370 &  & - & 0.7951 & 0.7582 & 0.2794 & 0.4723 & 0.3132 & 0.6110 \\
 & \cellcolor[HTML]{F2F2F2}APG & \cellcolor[HTML]{F2F2F2}0.8485 & \cellcolor[HTML]{F2F2F2}0.7848 & \cellcolor[HTML]{F2F2F2}0.2986 & \cellcolor[HTML]{F2F2F2}0.5123 & \cellcolor[HTML]{F2F2F2}0.3346 & \cellcolor[HTML]{F2F2F2}0.6478 &  & \cellcolor[HTML]{F2F2F2}APG & \cellcolor[HTML]{F2F2F2}0.8261 & \cellcolor[HTML]{F2F2F2}0.7650 & \cellcolor[HTML]{F2F2F2}0.2895 & \cellcolor[HTML]{F2F2F2}0.4891 & \cellcolor[HTML]{F2F2F2}0.3226 & \cellcolor[HTML]{F2F2F2}0.6202 \\
 & \cellcolor[HTML]{F2F2F2}Ours (APG) & \cellcolor[HTML]{F2F2F2}\textbf{0.8504} & \cellcolor[HTML]{F2F2F2}\textbf{0.7921} & \cellcolor[HTML]{F2F2F2}\textbf{0.3054} & \cellcolor[HTML]{F2F2F2}\textbf{0.5279} & \cellcolor[HTML]{F2F2F2}\textbf{0.3411} & \cellcolor[HTML]{F2F2F2}\textbf{0.6631} &  & \cellcolor[HTML]{F2F2F2}Ours (APG) & \cellcolor[HTML]{F2F2F2}\textbf{0.8386} & \cellcolor[HTML]{F2F2F2}\textbf{0.7846} & \cellcolor[HTML]{F2F2F2}\textbf{0.3058} & \cellcolor[HTML]{F2F2F2}\textbf{0.5179} & \cellcolor[HTML]{F2F2F2}\textbf{0.3412} & \cellcolor[HTML]{F2F2F2}\textbf{0.6568} \\
 & \cellcolor[HTML]{E7E6E6}DUET & \cellcolor[HTML]{E7E6E6}0.8454 & \cellcolor[HTML]{E7E6E6}0.7834 & \cellcolor[HTML]{E7E6E6}0.2861 & \cellcolor[HTML]{E7E6E6}0.5025 & \cellcolor[HTML]{E7E6E6}0.3238 & \cellcolor[HTML]{E7E6E6}0.6424 &  & \cellcolor[HTML]{E7E6E6}DUET & \cellcolor[HTML]{E7E6E6}0.8285 & \cellcolor[HTML]{E7E6E6}0.7686 & \cellcolor[HTML]{E7E6E6}0.2712 & \cellcolor[HTML]{E7E6E6}0.4750 & \cellcolor[HTML]{E7E6E6}0.3078 & \cellcolor[HTML]{E7E6E6}0.6191 \\
\multirow{-5}{*}{BERT4Rec} & \cellcolor[HTML]{E7E6E6}Ours (DUET) & \cellcolor[HTML]{E7E6E6}\textbf{0.8497} & \cellcolor[HTML]{E7E6E6}\textbf{0.7970} & \cellcolor[HTML]{E7E6E6}\textbf{0.3088} & \cellcolor[HTML]{E7E6E6}\textbf{0.5344} & \cellcolor[HTML]{E7E6E6}\textbf{0.3456} & \cellcolor[HTML]{E7E6E6}\textbf{0.6748} & \multirow{-5}{*}{BERT4Rec} & \cellcolor[HTML]{E7E6E6}Ours (DUET) & \cellcolor[HTML]{E7E6E6}\textbf{0.8326} & \cellcolor[HTML]{E7E6E6}\textbf{0.7811} & \cellcolor[HTML]{E7E6E6}\textbf{0.2992} & \cellcolor[HTML]{E7E6E6}\textbf{0.5104} & \cellcolor[HTML]{E7E6E6}\textbf{0.3329} & \cellcolor[HTML]{E7E6E6}\textbf{0.6435} \\
\midrule[1.2pt] \midrule[1.2pt]
\multicolumn{8}{c|}{\textbf{\texttt{Office}}} & \multicolumn{8}{c}{\textbf{\texttt{Scientific}}} \\
\midrule
 &  & \multicolumn{6}{c|}{\textbf{Metrics}} &  &  & \multicolumn{6}{c}{\textbf{Metrics}} \\ 
 \cline{3-8} \cline{11-16}
\multirow{-2}{*}{\textbf{SR Model}} & \multirow{-2}{*}{\textbf{DSR Model}} & AUC & UAUC & NDCG@10 & Recall@10 & NDCG@20 & Recall@20 & \multirow{-2}{*}{\textbf{SR Model}} & \multirow{-2}{*}{\textbf{DSR Model}} & AUC & UAUC & NDCG@10 & Recall@10 & NDCG@20 & Recall@20 \\
\midrule
 & - & 0.8158 & 0.7510 & 0.2701 & 0.4702 & 0.3046 & 0.6045 &  & - & 0.6100 & 0.5971 & 0.1337 & 0.2609 & 0.1648 & 0.3880 \\
 & \cellcolor[HTML]{F2F2F2}APG & \cellcolor[HTML]{F2F2F2}0.8359 & \cellcolor[HTML]{F2F2F2}0.7639 & \cellcolor[HTML]{F2F2F2}0.2862 & \cellcolor[HTML]{F2F2F2}0.4903 & \cellcolor[HTML]{F2F2F2}0.3202 & \cellcolor[HTML]{F2F2F2}0.6202 &  & \cellcolor[HTML]{F2F2F2}APG & \cellcolor[HTML]{F2F2F2}0.7310 & \cellcolor[HTML]{F2F2F2}0.6969 & \cellcolor[HTML]{F2F2F2}0.1700 & \cellcolor[HTML]{F2F2F2}0.3238 & \cellcolor[HTML]{F2F2F2}0.2099 & \cellcolor[HTML]{F2F2F2}0.4816 \\
 & \cellcolor[HTML]{F2F2F2}Ours (APG) & \cellcolor[HTML]{F2F2F2}\textbf{0.8394} & \cellcolor[HTML]{F2F2F2}\textbf{0.7673} & \cellcolor[HTML]{F2F2F2}\textbf{0.2764} & \cellcolor[HTML]{F2F2F2}\textbf{0.4823} & \cellcolor[HTML]{F2F2F2}\textbf{0.3128} & \cellcolor[HTML]{F2F2F2}\textbf{0.6222} &  & \cellcolor[HTML]{F2F2F2}Ours (APG) & \cellcolor[HTML]{F2F2F2}\textbf{0.7315} & \cellcolor[HTML]{F2F2F2}\textbf{0.6989} & \cellcolor[HTML]{F2F2F2}\textbf{0.1746} & \cellcolor[HTML]{F2F2F2}\textbf{0.3429} & \cellcolor[HTML]{F2F2F2}\textbf{0.2147} & \cellcolor[HTML]{F2F2F2}\textbf{0.5020} \\
 & \cellcolor[HTML]{E7E6E6}DUET & \cellcolor[HTML]{E7E6E6}0.8297 & \cellcolor[HTML]{E7E6E6}0.7531 & \cellcolor[HTML]{E7E6E6}0.2813 & \cellcolor[HTML]{E7E6E6}0.4816 & \cellcolor[HTML]{E7E6E6}0.3147 & \cellcolor[HTML]{E7E6E6}0.6085 &  & \cellcolor[HTML]{E7E6E6}DUET & \cellcolor[HTML]{E7E6E6}0.6714 & \cellcolor[HTML]{E7E6E6}0.6266 & \cellcolor[HTML]{E7E6E6}0.1428 & \cellcolor[HTML]{E7E6E6}0.2736 & \cellcolor[HTML]{E7E6E6}0.1748 & \cellcolor[HTML]{E7E6E6}0.3979 \\
\multirow{-5}{*}{DIN} & \cellcolor[HTML]{E7E6E6}Ours (DUET) & \cellcolor[HTML]{E7E6E6}\textbf{0.8361} & \cellcolor[HTML]{E7E6E6}\textbf{0.7642} & \cellcolor[HTML]{E7E6E6}\textbf{0.2949} & \cellcolor[HTML]{E7E6E6}\textbf{0.4970} & \cellcolor[HTML]{E7E6E6}\textbf{0.3282} & \cellcolor[HTML]{E7E6E6}\textbf{0.6240} & \multirow{-5}{*}{DIN} & \cellcolor[HTML]{E7E6E6}Ours (DUET) & \cellcolor[HTML]{E7E6E6}\textbf{0.7138} & \cellcolor[HTML]{E7E6E6}\textbf{0.6682} & \cellcolor[HTML]{E7E6E6}\textbf{0.1589} & \cellcolor[HTML]{E7E6E6}\textbf{0.3012} & \cellcolor[HTML]{E7E6E6}\textbf{0.1989} & \cellcolor[HTML]{E7E6E6}\textbf{0.4573} \\
\midrule
 & - & 0.8346 & 0.7606 & 0.2704 & 0.4762 & 0.3055 & 0.6117 &  & - & 0.7424 & 0.7094 & 0.1621 & 0.3214 & 0.2049 & 0.4952 \\
 & \cellcolor[HTML]{F2F2F2}APG & \cellcolor[HTML]{F2F2F2}0.8343 & \cellcolor[HTML]{F2F2F2}0.7623 & \cellcolor[HTML]{F2F2F2}0.2809 & \cellcolor[HTML]{F2F2F2}0.4831 & \cellcolor[HTML]{F2F2F2}0.3154 & \cellcolor[HTML]{F2F2F2}0.6159 &  & \cellcolor[HTML]{F2F2F2}APG & \cellcolor[HTML]{F2F2F2}0.7273 & \cellcolor[HTML]{F2F2F2}0.6933 & \cellcolor[HTML]{F2F2F2}0.1592 & \cellcolor[HTML]{F2F2F2}0.3159 & \cellcolor[HTML]{F2F2F2}0.1988 & \cellcolor[HTML]{F2F2F2}0.4758 \\
 & \cellcolor[HTML]{F2F2F2}Ours (APG) & \cellcolor[HTML]{F2F2F2}\textbf{0.8354} & \cellcolor[HTML]{F2F2F2}\textbf{0.7671} & \cellcolor[HTML]{F2F2F2}\textbf{0.2914} & \cellcolor[HTML]{F2F2F2}\textbf{0.4966} & \cellcolor[HTML]{F2F2F2}\textbf{0.3255} & \cellcolor[HTML]{F2F2F2}\textbf{0.6272} &  & \cellcolor[HTML]{F2F2F2}Ours (APG) & \cellcolor[HTML]{F2F2F2}\textbf{0.7402} & \cellcolor[HTML]{F2F2F2}\textbf{0.7133} & \cellcolor[HTML]{F2F2F2}\textbf{0.1859} & \cellcolor[HTML]{F2F2F2}\textbf{0.3535} & \cellcolor[HTML]{F2F2F2}\textbf{0.2273} & \cellcolor[HTML]{F2F2F2}\textbf{0.5161} \\
 & \cellcolor[HTML]{E7E6E6}DUET & \cellcolor[HTML]{E7E6E6}0.8399 & \cellcolor[HTML]{E7E6E6}0.7649 & \cellcolor[HTML]{E7E6E6}0.2930 & \cellcolor[HTML]{E7E6E6}0.4976 & \cellcolor[HTML]{E7E6E6}0.3268 & \cellcolor[HTML]{E7E6E6}0.6262 &  & \cellcolor[HTML]{E7E6E6}DUET & \cellcolor[HTML]{E7E6E6}0.7270 & \cellcolor[HTML]{E7E6E6}0.6881 & \cellcolor[HTML]{E7E6E6}0.1658 & \cellcolor[HTML]{E7E6E6}0.3224 & \cellcolor[HTML]{E7E6E6}0.2036 & \cellcolor[HTML]{E7E6E6}0.4703 \\
\multirow{-5}{*}{GRU4Rec} & \cellcolor[HTML]{E7E6E6}Ours (DUET) & \cellcolor[HTML]{E7E6E6}\textbf{0.8437} & \cellcolor[HTML]{E7E6E6}\textbf{0.7737} & \cellcolor[HTML]{E7E6E6}\textbf{0.3072} & \cellcolor[HTML]{E7E6E6}\textbf{0.5112} & \cellcolor[HTML]{E7E6E6}\textbf{0.3403} & \cellcolor[HTML]{E7E6E6}\textbf{0.6366} & \multirow{-5}{*}{GRU4Rec} & \cellcolor[HTML]{E7E6E6}Ours (DUET) & \cellcolor[HTML]{E7E6E6}\textbf{0.7410} & \cellcolor[HTML]{E7E6E6}\textbf{0.7054} & \cellcolor[HTML]{E7E6E6}\textbf{0.1792} & \cellcolor[HTML]{E7E6E6}\textbf{0.3415} & \cellcolor[HTML]{E7E6E6}\textbf{0.2196} & \cellcolor[HTML]{E7E6E6}\textbf{0.5020} \\
\midrule
 & - & 0.8288 & 0.7587 & 0.2820 & 0.4858 & 0.3153 & 0.6151 &  & - & 0.7175 & 0.6772 & 0.1587 & 0.3145 & 0.1960 & 0.4631 \\
 & \cellcolor[HTML]{F2F2F2}APG & \cellcolor[HTML]{F2F2F2}0.8377 & \cellcolor[HTML]{F2F2F2}0.7603 & \cellcolor[HTML]{F2F2F2}0.2823 & \cellcolor[HTML]{F2F2F2}0.4804 & \cellcolor[HTML]{F2F2F2}0.3170 & \cellcolor[HTML]{F2F2F2}0.6117 &  & \cellcolor[HTML]{F2F2F2}APG & \cellcolor[HTML]{F2F2F2}0.6952 & \cellcolor[HTML]{F2F2F2}0.6610 & \cellcolor[HTML]{F2F2F2}0.1523 & \cellcolor[HTML]{F2F2F2}0.3040 & \cellcolor[HTML]{F2F2F2}0.1910 & \cellcolor[HTML]{F2F2F2}0.4583 \\
 & \cellcolor[HTML]{F2F2F2}Ours (APG) & \cellcolor[HTML]{F2F2F2}\textbf{0.8402} & \cellcolor[HTML]{F2F2F2}\textbf{0.7679} & \cellcolor[HTML]{F2F2F2}\textbf{0.2997} & \cellcolor[HTML]{F2F2F2}\textbf{0.4995} & \cellcolor[HTML]{F2F2F2}\textbf{0.3333} & \cellcolor[HTML]{F2F2F2}\textbf{0.6269} &  & \cellcolor[HTML]{F2F2F2}Ours (APG) & \cellcolor[HTML]{F2F2F2}\textbf{0.7161} & \cellcolor[HTML]{F2F2F2}\textbf{0.6728} & \cellcolor[HTML]{F2F2F2}\textbf{0.1634} & \cellcolor[HTML]{F2F2F2}\textbf{0.3122} & \cellcolor[HTML]{F2F2F2}\textbf{0.2002} & \cellcolor[HTML]{F2F2F2}\textbf{0.4580} \\
 & \cellcolor[HTML]{E7E6E6}DUET & \cellcolor[HTML]{E7E6E6}0.8395 & \cellcolor[HTML]{E7E6E6}0.7594 & \cellcolor[HTML]{E7E6E6}0.2833 & \cellcolor[HTML]{E7E6E6}0.4831 & \cellcolor[HTML]{E7E6E6}0.3173 & \cellcolor[HTML]{E7E6E6}0.6105 &  & \cellcolor[HTML]{E7E6E6}DUET & \cellcolor[HTML]{E7E6E6}0.6992 & \cellcolor[HTML]{E7E6E6}0.6565 & \cellcolor[HTML]{E7E6E6}0.1579 & \cellcolor[HTML]{E7E6E6}0.3040 & \cellcolor[HTML]{E7E6E6}0.1944 & \cellcolor[HTML]{E7E6E6}0.4481 \\
\multirow{-5}{*}{SASRec} & \cellcolor[HTML]{E7E6E6}Ours (DUET) & \cellcolor[HTML]{E7E6E6}\textbf{0.8460} & \cellcolor[HTML]{E7E6E6}\textbf{0.7735} & \cellcolor[HTML]{E7E6E6}\textbf{0.2997} & \cellcolor[HTML]{E7E6E6}\textbf{0.5061} & \cellcolor[HTML]{E7E6E6}\textbf{0.3345} & \cellcolor[HTML]{E7E6E6}\textbf{0.6380} & \multirow{-5}{*}{SASRec} & \cellcolor[HTML]{E7E6E6}Ours (DUET) & \cellcolor[HTML]{E7E6E6}\textbf{0.7111} & \cellcolor[HTML]{E7E6E6}\textbf{0.6738} & \cellcolor[HTML]{E7E6E6}\textbf{0.1548} & \cellcolor[HTML]{E7E6E6}\textbf{0.3016} & \cellcolor[HTML]{E7E6E6}\textbf{0.1957} & \cellcolor[HTML]{E7E6E6}\textbf{0.4614} \\
\midrule
 & - & 0.8184 & 0.7544 & 0.2701 & 0.4732 & 0.3049 & 0.6092 &  & - & 0.7329 & 0.7000 & 0.1744 & 0.3306 & 0.2108 & 0.4768 \\
 & \cellcolor[HTML]{F2F2F2}APG & \cellcolor[HTML]{F2F2F2}0.8354 & \cellcolor[HTML]{F2F2F2}0.7633 & \cellcolor[HTML]{F2F2F2}0.2885 & \cellcolor[HTML]{F2F2F2}0.4923 & \cellcolor[HTML]{F2F2F2}0.3223 & \cellcolor[HTML]{F2F2F2}0.6222 &  & \cellcolor[HTML]{F2F2F2}APG & \cellcolor[HTML]{F2F2F2}0.7255 & \cellcolor[HTML]{F2F2F2}0.6953 & \cellcolor[HTML]{F2F2F2}0.1699 & \cellcolor[HTML]{F2F2F2}0.3306 & \cellcolor[HTML]{F2F2F2}0.2069 & \cellcolor[HTML]{F2F2F2}0.4758 \\
 & \cellcolor[HTML]{F2F2F2}Ours (APG) & \cellcolor[HTML]{F2F2F2}\textbf{0.8462} & \cellcolor[HTML]{F2F2F2}\textbf{0.7767} & \cellcolor[HTML]{F2F2F2}\textbf{0.3032} & \cellcolor[HTML]{F2F2F2}\textbf{0.5130} & \cellcolor[HTML]{F2F2F2}\textbf{0.3374} & \cellcolor[HTML]{F2F2F2}\textbf{0.6419} &  & \cellcolor[HTML]{F2F2F2}Ours (APG) & \cellcolor[HTML]{F2F2F2}\textbf{0.7456} & \cellcolor[HTML]{F2F2F2}\textbf{0.7132} & \cellcolor[HTML]{F2F2F2}\textbf{0.1760} & \cellcolor[HTML]{F2F2F2}\textbf{0.3508} & \cellcolor[HTML]{F2F2F2}\textbf{0.2183} & \cellcolor[HTML]{F2F2F2}\textbf{0.5167} \\
 & \cellcolor[HTML]{E7E6E6}DUET & \cellcolor[HTML]{E7E6E6}0.8371 & \cellcolor[HTML]{E7E6E6}0.7682 & \cellcolor[HTML]{E7E6E6}0.2842 & \cellcolor[HTML]{E7E6E6}0.4900 & \cellcolor[HTML]{E7E6E6}0.3187 & \cellcolor[HTML]{E7E6E6}0.6223 &  & \cellcolor[HTML]{E7E6E6}DUET & \cellcolor[HTML]{E7E6E6}0.7325 & \cellcolor[HTML]{E7E6E6}0.6962 & \cellcolor[HTML]{E7E6E6}0.1707 & \cellcolor[HTML]{E7E6E6}0.3262 & \cellcolor[HTML]{E7E6E6}0.2090 & \cellcolor[HTML]{E7E6E6}0.4785 \\
\multirow{-5}{*}{BERT4Rec} & \cellcolor[HTML]{E7E6E6}Ours (DUET) & \cellcolor[HTML]{E7E6E6}\textbf{0.8380} & \cellcolor[HTML]{E7E6E6}\textbf{0.7731} & \cellcolor[HTML]{E7E6E6}\textbf{0.2892} & \cellcolor[HTML]{E7E6E6}\textbf{0.4987} & \cellcolor[HTML]{E7E6E6}\textbf{0.3249} & \cellcolor[HTML]{E7E6E6}\textbf{0.6365} & \multirow{-5}{*}{BERT4Rec} & \cellcolor[HTML]{E7E6E6}Ours (DUET) & \cellcolor[HTML]{E7E6E6}\textbf{0.7420} & \cellcolor[HTML]{E7E6E6}\textbf{0.7108} & \cellcolor[HTML]{E7E6E6}\textbf{0.1826} & \cellcolor[HTML]{E7E6E6}\textbf{0.3477} & \cellcolor[HTML]{E7E6E6}\textbf{0.2235} & \cellcolor[HTML]{E7E6E6}\textbf{0.5099} \\
\bottomrule[2pt]
\end{tabular}
}
\vspace{-0.3cm}
\end{table*}
\section{Experiments}
\label{sec:experiments}
\subsection{Experimental Setup}
\subsubsection{Datasets and Preprocessing.}
\begin{sloppypar}
We evaluate SOLID and baselines on eight datasets. \texttt{Amazon Arts (Arts)}, \texttt{Amazon Instruments (Instruments)}, \texttt{Amazon Office (Office)}, \texttt{Amazon Scientific (Scientific)}, which are four benchmarks that was recently released but has been widely used in the multimodal recommendation tasks~\cite{ref:wang2023missrec}.
\texttt{Amazon CDs (CDs)}, \texttt{Amazon Electronic (Electronic)}, \texttt{Douban Book (Book)}, and \texttt{Douban Music (Music)}, which are four widely used public benchmarks in the recommendation tasks. 
We choose the leave-one-out approach to process the dataset, taking the last action of each user for testing and all previous actions for training and validation. Our task is CTR (Click-through Rate) prediction, so we process these datasets into CTR prediction datasets. These datasets consist of user rating datasets with complete reviews. We treat all user-item interactions in the dataset as positive samples because having a rating implies that the user clicked on the item. Further, to ensure the training process goes smoothly with both positive and negative samples, we sample $4$ negative samples for each positive sample in the training set and $99$ negative samples for each positive sample in the test set.
\end{sloppypar}

\subsubsection{Baselines.}
\label{subsec:experiment_baseline}

The baselines we select are as follows:
\begin{itemize}[itemsep=2pt,topsep=2pt,leftmargin=20pt]
\item \textbf{{Static Recommendation Models.}}
    \textit{DIN}~\cite{ref:din}, \textit{GRU4Rec}~\cite{ref:gru4rec}, \textit{SASRec}~\cite{ref:sasrec}, and \textit{BERT4Rec}~\cite{ref:bert4rec} are all highly prevalent sequential recommendation methods in both academic research and the industry. They each incorporate different techniques, such as Attention, GRU (Gated Recurrent Unit), and Self-Attention, to enhance the recommendation process. 

\begin{sloppypar}
\item \textbf{{Dynamic Recommendation Models.}}
    \textit{DUET}~\cite{ref:duet} and \textit{APG}~\cite{ref:apg_rs1} consists of two parts: a parameter generation model and a primary model. The primary model refers to the aforementioned models like DIN, GRU4Rec, SASRec, BERT4Rec, etc. After pre-training, the parameter generation model can generate model parameters for the primary model during inference based on the samples.
\end{sloppypar}

\end{itemize}

\subsubsection{Evaluation Metrics} 
We use the widely adopted \textit{AUC}, \textit{UAUC}, \textit{NDCG}, and \textit{Recall} as the metrics to evaluate model performance. 

\subsection{Overall Results}
\begin{figure}[!h]
    \centering
    \includegraphics[width=0.45\textwidth]{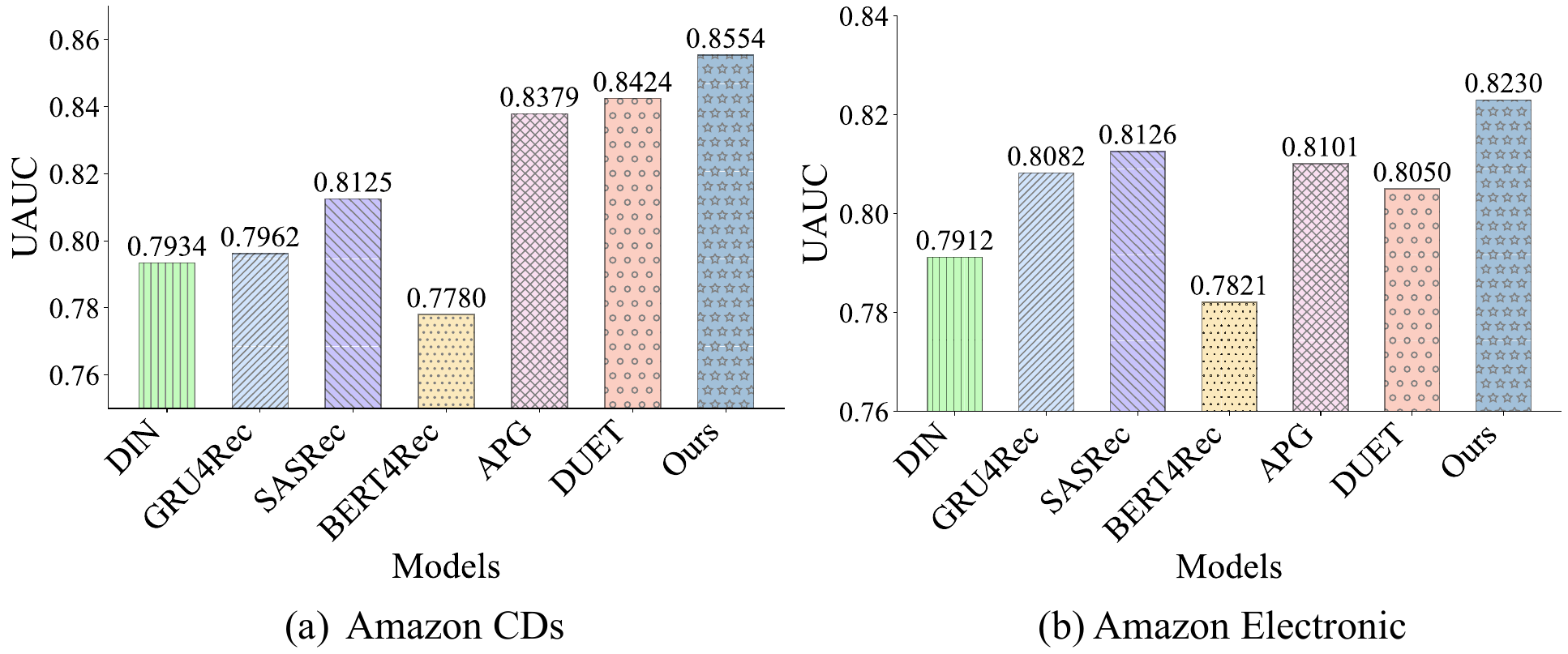}
    \vspace{-0.3cm}
    \caption{UAUC comparison of the proposed method and baseline on the \texttt{CDs} and \texttt{Electronic} datasets.}
    \label{fig:main_fig_douban}
    \vspace{-0.3cm}
\end{figure}

\begin{figure}[!h]
    \centering
    \includegraphics[width=0.45\textwidth]{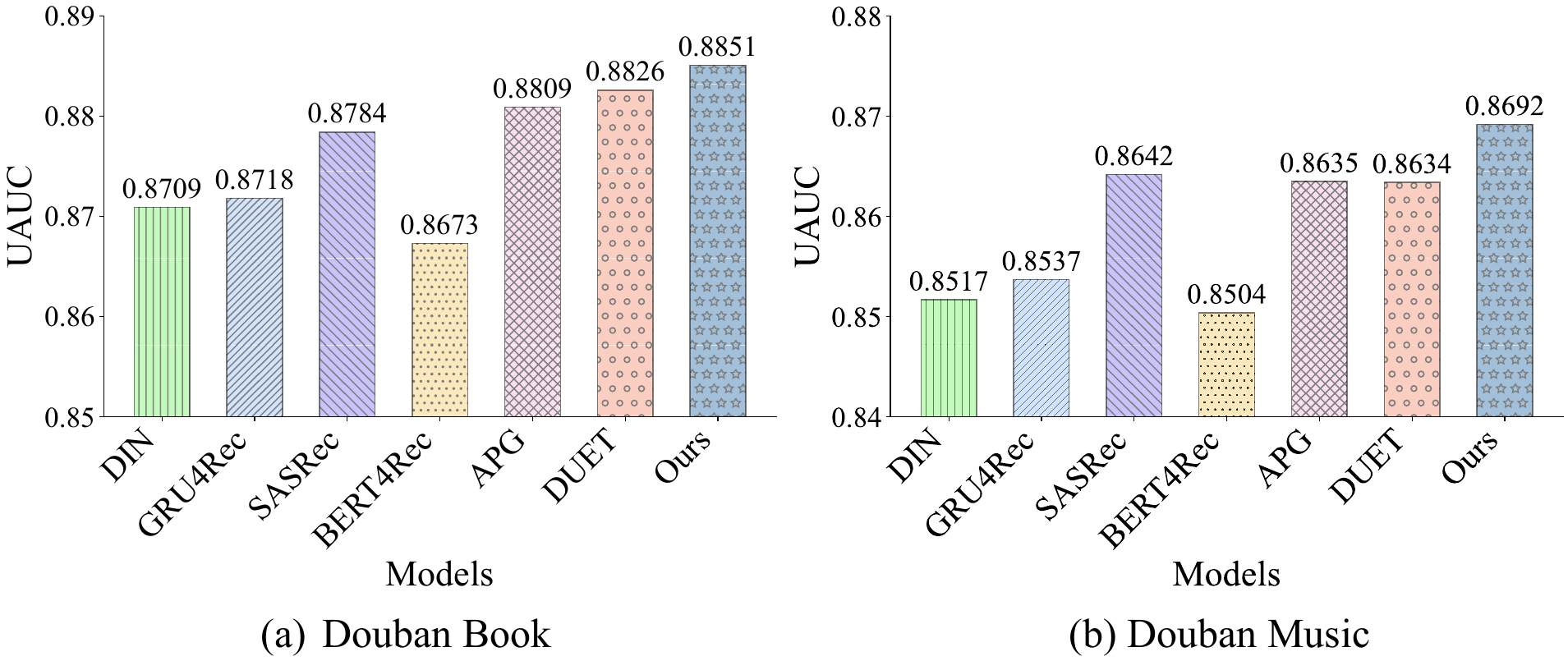}
    \vspace{-0.3cm}
    \caption{UAUC comparison of the proposed method and baseline on the \texttt{Book} and \texttt{Music} datasets.}
    \label{fig:main_fig_amazon}
    \vspace{-0.3cm}
\end{figure}

As shown in Table~\ref{tab:main}, we evaluate the overall performance across four multimodal datasets: Arts, Instruments, Office, and Scientific. For each dataset, we test the performance of four SR Models: DIN, GRU4Rec, SASRec, and BERT4Rec. 
We evaluate performance via AUC, UAUC, NDCG@10, Recall@10, NDCG@20, and Recall@20. 
For each SR Model, there are five options for DSR Models: None (``-''), APG, Ours (APG), DUET, and Ours (DUET), where ``-'' indicates no DSR Model usage, i.e., the inherent performance of the SR Model itself. Since the ``-'' option consistently performs worse than using a DSR Model, our comparison primarily focuses on the performance of APG vs. Ours (APG) and DUET vs. Ours (DUET) for each SR Model. Across all datasets, all SR Models, and all metrics, our proposed methods significantly outperform both APG and DUET.
We conducted experiments on four other commonly used recommendation datasets and compared the UAUC metric in Figures ~\ref{fig:main_fig_douban} and \ref{fig:main_fig_amazon}.
Our method (\{SR=SASRec, DSR=DUET\}) significantly outperforms other SR and DSR Models across all the datasets.

\subsection{Ablation Study}
We conduct ablation studies on each dataset, each SR and each DSR to further analyze the impact of modules and modalities. The ablation results on each dataset, DR, and DSR combinations are similar, so we only show the results under the condition \{Dataset=Arts, SR=SASRec, DSR=DUET\}.
Each row's \Checkmark and \XSolidBrush respectively indicate with and without the module/modality.
\subsubsection{Ablation Study on Modules.}
As shown in Table \ref{tab:ablation_module}, we conduct an ablation study on each module proposed in our method, SPG stands for Semantic Parameter Generation, SML stands for Semantic Metacode Learning, and SCL stands for Semancic Codebook Learning. Since SPG is a prerequisite for SML, SML cannot exist independently of SPG; therefore, there is no separate performance data for SML alone in the table. 
The first line represents the traditional DSR model where parameters are generated using an item sequence. The second line represents generating parameters using a semantic sequence. The third line represents the joint generation of parameters using both item sequence and semantic sequence, with joint training. The fourth line represents using semantic codebook learning without using semantic information. The fifth line represents our complete method.
The experiments show that the model performs best when all three modules are used. In terms of individual modules, SCL has the greatest impact on performance.
\begin{table}[ht]
\vspace{-0.1cm}
\caption{Results of the ablation study over our proposed methods with respect to the modules. The \textbf{best} results is  in bold.}
\label{tab:ablation_module}
\vspace{-0.3cm}
\renewcommand{\arraystretch}{1.02}
\resizebox{0.475\textwidth}{!}{
\begin{tabular}{c|c|c|c|c|c|c|c|c}
\toprule[2pt]
\multicolumn{3}{c|}{\textbf{Module}} & \multicolumn{6}{c}{\textbf{Metrics}} \\
\midrule
SPG & SML & SCL & AUC & UAUC & NDCG@10 & Recall@10 & NDCG@20 & Recall@20 \\
\midrule \midrule
\XSolidBrush & \XSolidBrush & \XSolidBrush & 0.8345 & 0.7660 & 0.2727 & 0.4763 & 0.3101 & 0.6177 \\
\Checkmark & \XSolidBrush & \XSolidBrush & 0.8459 & 0.7783 & 0.2905 & 0.5069 & 0.3270 & 0.6425 \\
\Checkmark & \Checkmark & \XSolidBrush & 0.8270 & 0.7530 & 0.2491 & 0.4539 & 0.2857 & 0.5922 \\
\XSolidBrush & \XSolidBrush & \Checkmark & 0.8461 & 0.7828 & 0.2979 & 0.5166 & 0.3326 & 0.6481 \\
\Checkmark & \Checkmark & \Checkmark & \textbf{0.8469} & \textbf{0.7867} & \textbf{0.3022} & \textbf{0.5216} & \textbf{0.3382} & \textbf{0.6560} \\
\bottomrule[2pt]
\end{tabular}
}
\vspace{-0.3cm}
\end{table}

\subsubsection{Ablation Study on Modalities.}
As shown in Table \ref{tab:ablation_modal}, we conduct ablation study on each modality. The experimental results show that the fusion of three modalities—ID, Image, and Text—is not necessarily the best option. In terms of the impact on performance for individual modalities, Text > Image > ID. For the fusion of two modalities, in terms of impact on performance, ID + Text > Image + Text > ID + Image.
\begin{table}[ht]
\vspace{-0.1cm}
\caption{Results of the ablation study over our proposed methods with respect to the modalities. The \textbf{best} results is  in bold.}
\label{tab:ablation_modal}
\vspace{-0.3cm}
\renewcommand{\arraystretch}{1.02}
\resizebox{0.475\textwidth}{!}{
\begin{tabular}{c|c|c|c|c|c|c|c|c}
\toprule[2pt]
\multicolumn{3}{c|}{\textbf{Modality}} & \multicolumn{6}{c}{\textbf{Metrics}} \\
\midrule
\multicolumn{1}{c|}{ID} & \multicolumn{1}{c|}{Image} & \multicolumn{1}{c|}{Text} & \multicolumn{1}{c|}{AUC} & \multicolumn{1}{c|}{UAUC} & \multicolumn{1}{c|}{NDCG@10} & \multicolumn{1}{c|}{Recall@10} & \multicolumn{1}{c|}{NDCG@20} & \multicolumn{1}{c}{Recall@20} \\
\midrule \midrule
\Checkmark & \XSolidBrush & \XSolidBrush & 0.8479 & 0.7850 & 0.2983 & 0.5155 & 0.3347 & 0.6510  \\
\XSolidBrush & \Checkmark & \XSolidBrush & 0.8438 & 0.7818 & 0.2953 & 0.5117 & 0.3310 & 0.6476  \\
\XSolidBrush & \XSolidBrush & \Checkmark & {\ul 0.8480} & 0.7858 & \textbf{0.3031} & \textbf{0.5252} & {\ul 0.3379} & 0.6548 \\
\Checkmark & \Checkmark & \XSolidBrush & 0.8459 & 0.7832 & 0.2953 & 0.5148 & 0.3313 & 0.6492 \\
\Checkmark & \XSolidBrush & \Checkmark & \textbf{0.8490} & \textbf{0.7881} & 0.3016 & {\ul 0.5223} & 0.3376 & \textbf{0.6566} \\
\XSolidBrush & \Checkmark & \Checkmark & 0.8471 & 0.7857 & 0.2963 & 0.5173 & 0.3319 & 0.6513 \\
\Checkmark & \Checkmark & \Checkmark & 0.8469 & {\ul 0.7867} & {\ul 0.3022} & 0.5216 & \textbf{0.3382} & {\ul 0.6560} \\
\bottomrule[2pt]
\end{tabular}
}
\vspace{-0.3cm}
\end{table}

\subsection{Depth Analysis}
We further conduct depth analysis to demonstrate the effectiveness. Unless otherwise specified, the dataset, SR, and DSR default to Arts, SASRec, and DUET, respectively. Note that we get similar results for all settings, but only a subset of them are shown here.
\subsubsection{Stability and Robustness}
We tested the variance of the UAUC for SOLID and DUET on each user in the Arts dataset when faced with similar user behaviors. Specifically, we added one user behavior at a time for each user behavior and calculated the performance variance. We then aggregated the variances for all users to obtain the median, mean, minimum, and maximum of these variances. Table~\ref{tab:var_compare} shows that SOLID has stronger stability and robustness compared to DUET.
\begin{table}[!h]
\vspace{-0.1cm}
\centering
\caption{Variance comparison.}
\label{tab:var_compare}
\vspace{-0.3cm}
\renewcommand{\arraystretch}{1.02}
\resizebox{0.38\textwidth}{!}{
\begin{tabular}{c|c|c|c|c|c|c|c}
\bottomrule[2pt]
\multicolumn{4}{c|}{\textbf{DUET}} & \multicolumn{4}{c}{\textbf{Ours}} \\
\hline
Medium & Mean & Min & Max & Medium & Mean & Min & Max \\
\hline
0.35 & 0.42 & 0.08 & 0.69 & 0.26 & 0.29 & 0.03 & 0.47 \\
\bottomrule[2pt]
\end{tabular}
}
\vspace{-0.3cm}
\end{table}

\subsubsection{Cost Comparison}
In Table~\ref{tab:cost}, we do analysis based on the BERT4Rec (the biggest SR in our paper), the increased memory and time are not important because the increase is slight and does not affect real-time performance~\cite{ref:duet,ref:device_cloud}.
\begin{table}[!h]
\vspace{-0.1cm}
\caption{Cost of our method.}
\label{tab:cost}
\vspace{-0.3cm}
\renewcommand{\arraystretch}{1.02}
\resizebox{0.47\textwidth}{!}{
\begin{tabular}{c|c|c|c|c|c}
\bottomrule[2pt]
\multicolumn{3}{c|}{\textbf{DUET}} & \multicolumn{3}{c}{\textbf{Ours}} \\
\hline
\#Param. & Train (s/epoch) & Test (s/batch) & \#Param. & Train (s/epoch) & Test (s/batch) \\
\hline
695.84k & 106.0106 & 0.0084 & 821.44k & 130.6742 & 0.0103 \\
\bottomrule[2pt]
\end{tabular}
}
\vspace{-0.3cm}
\end{table}

\subsubsection{Hyperparameter Analysis}

\begin{figure}[ht]
\vspace{-0.1cm}
  \centering
  \includegraphics[width=0.78\linewidth]{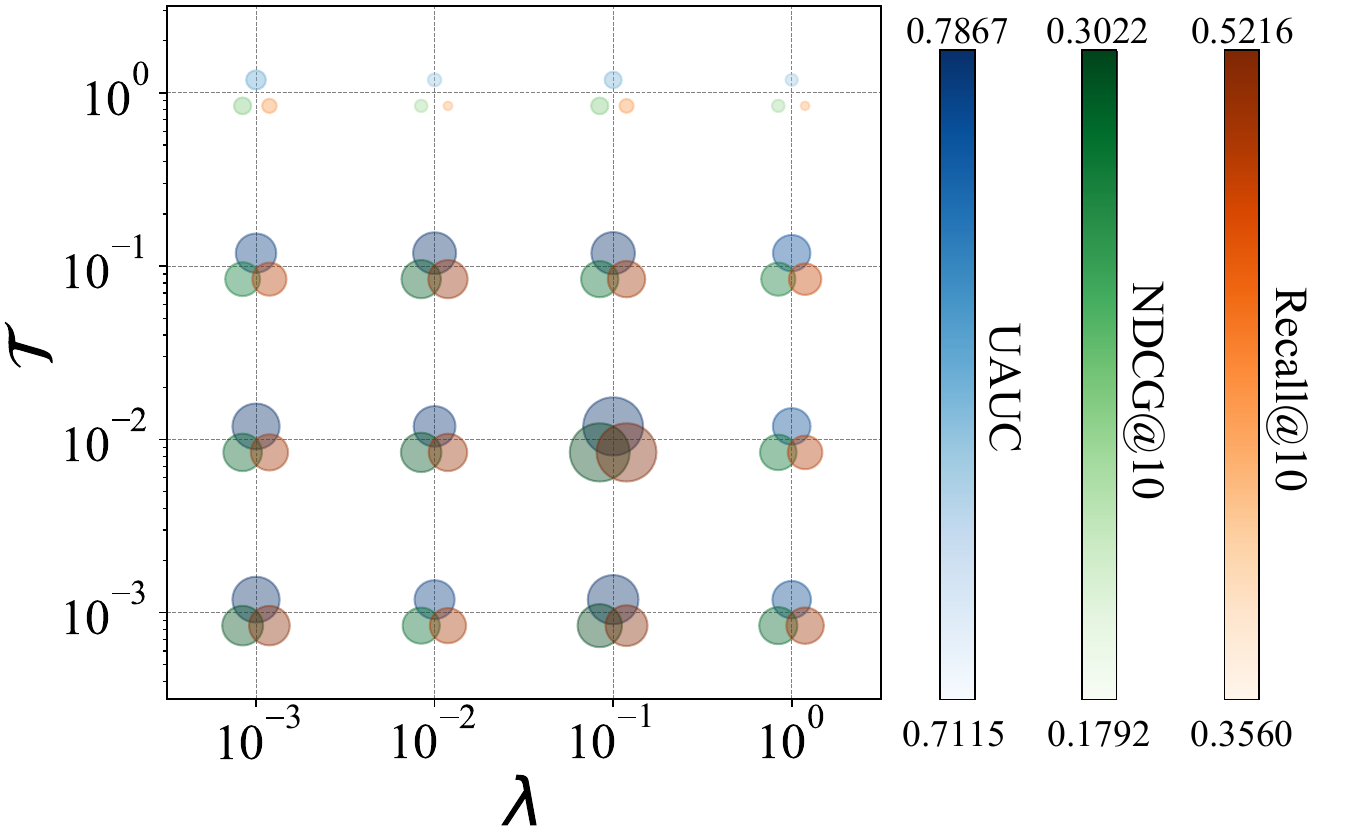}
  \vspace{-0.3cm}
  \caption{Hyperparameter Grid Search.}
  \label{fig:hyperparam_grid_search}
\vspace{-0.3cm}
\end{figure}

To analyze the impace of the main hyperparameters \(\lambda\) and \(\mathcal{T}\), we conduct grid search experiment.
As shown in Figure \ref{fig:hyperparam_grid_search}, the horizontal axis represents \(\lambda\), and the vertical axis represents \(\mathcal{T}\). The depth of the color and the radius of the circle represent the magnitude of the value; the larger the value, the deeper the color and the larger the circle (i.e., the larger the radius). Blue, green, and orange represent the metrics UAUC, NDCG@10, and Recall@10, respectively.
The results show that the best performance is achieved when \(\lambda=0.1\) and \(\mathcal{T}=0.01\).
\section{Conclusion}
\label{sec:conclusion}
In this paper, we have presented the Semantic Codebook Learning for Dynamic Recommendation Models (SOLID) as a solution to the limitations faced by existing dynamic sequence recommendation systems (DSR). Our framework integrates multimodal information, including images and text, with user-item interactions to enhance recommendation accuracy and adaptability. By disentangling model parameters into trunk parameters capturing generalized user behavior trends and branch parameters tailored to individual user actions, SOLID offers a more efficient and effective recommendation system. Through extensive experimentation across multiple datasets, we have demonstrated that SOLID significantly outperforms previous DSR models, with an significant improvement on extensive datasets and models. These results underscore the potential of leveraging multimodal information to advance the capabilities of dynamic recommendation systems, paving the way for more personalized and responsive user experiences in the era of digital personalization.
\clearpage
\section*{ACKNOWLEDGEMENTS}
\begin{sloppypar}
National Science and Technology Major Project (2022ZD0119100), Scientific Research Fund of Zhejiang Provincial Education Department (Y202353679), Key Research and Development Program of Zhejiang Province (No. 2024C03270), National Natural Science Foundation of China (No. 62441605), the StarryNight Science Fund of Zhejiang University Shanghai Institute for Advanced Study (SN-ZJU-SIAS-0010).
\end{sloppypar}
\bibliographystyle{ACM-Reference-Format}
\bibliography{reference}
\appendix

\section{Appendix}
\label{sec:appendix}
This is the Appendix for ``Semantic Codebook Learning for Dynamic Recommendation Models''.

\subsection{Supplementary Experiments}
\subsubsection{Datasets.}
The statistics of the datasets used in the experiments is shown in Table~\ref{tab:statistics_of_datasets}.
\begin{table}[!h]
\vspace{-0.2cm}
\caption{Statistics of Datasets.} 
\label{tab:statistics_of_datasets}
\vspace{-0.3cm}
\begin{tabular}{c|c|c|c|c}
\toprule[2pt]
\textbf{Dataset} & \textbf{\#User} & \textbf{\#Item} & \textbf{\#Interaction} & \textbf{Density} \\
\midrule \midrule
Arts & 45,486 & 21,019 & 395,150 & 0.0004133 \\
\rowcolor[HTML]{F2F2F2} 
Office & 87,436 & 25,986 & 684,837 & 0.0003014 \\
Instruments & 24,962 & 9,964 & 208,926 & 0.0008400 \\
\rowcolor[HTML]{F2F2F2} 
Scientific & 8,442 & 4,385 & 59,427 & 0.0016053 \\
CDs & 1,578,597 & 486,360 & 3,749,004 & 0.0000049 \\
\rowcolor[HTML]{F2F2F2} 
Electronic & 4,201,696 & 476,002 & 7,824,482 & 0.0000039 \\
Book & 46,549 & 212,996 & 1,861,533 & 0.0001878 \\
\rowcolor[HTML]{F2F2F2} 
Music & 39,743 & 164,224 & 1,792,502 & 0.0002746 \\
\bottomrule[2pt]
\end{tabular}
\end{table}

\subsubsection{Hyperparameters and Training Schedules}
\label{sec:appendix_implementation_detail}

We summarize the hyperparameters and training schedules of the datasets used in the experiments in Table~\ref{tab:hyperparameters_and_training_schedule}.

\begin{table}[!h]
\vspace{-0.1cm}
    \caption{Hyperparameters and training schedules of SOLID.}
    \label{tab:hyperparameters_and_training_schedule}
    \centering
    \vspace{-0.3cm}
 \resizebox{0.43\textwidth}{!}{
    \begin{tabular}{c|c|c}
    \toprule[2pt]
    \textbf{Dataset} & \textbf{Parameters} & \textbf{Setting} \\ 
    \midrule[1pt]
    \midrule[1pt]
    \multirow{11}{*}{\makecell[c]{Arts\\Office\\Instruments\\Scientific\\CDs\\Electronic\\Book\\Music}} & GPU & Tesla A100 \\ \cline{2-3}
    \multirow{11}{*}{} & Optimizer & Adam\\ \cline{2-3}
    \multirow{11}{*}{} & \makecell[c]{Learning Rate} & 0.001\\ \cline{2-3}
    \multirow{11}{*}{} & \makecell[c]{Batch Size} & 1024 \\ \cline{2-3}
    \multirow{11}{*}{} & \makecell[c]{Sequence Length} & 10 \\ \cline{2-3}
    \multirow{11}{*}{} & \makecell[c]{the Dimension of Embedding} & 1×32 \\  \cline{2-3}
    \multirow{11}{*}{} & \makecell[c]{the Amount of MLP} & 2 \\  \cline{2-3}
    \multirow{11}{*}{} & \makecell[c]{Hidden Dimension of \\ Semantic Codebook} & 64 \\
    \cline{2-3}
    \multirow{11}{*}{} & \makecell[c]{$\mathbf{z}$ Dimension of \\ Semantic Codebook} & 32 \\
     \bottomrule[2pt]
    \end{tabular}
   }
\end{table}

\end{document}